\documentclass[lineno]{biometrika}

\usepackage{amsmath}

\usepackage{times}
\usepackage{bm}
\usepackage{natbib}

\usepackage[plain,noend]{algorithm2e}

\makeatletter
\renewcommand{\algocf@captiontext}[2]{#1\algocf@typo. \AlCapFnt{}#2} 
\def\@algocf@capt@plain{top}
\renewcommand{\algocf@makecaption}[2]{%
  \addtolength{\hsize}{\algomargin}%
  \sbox\@tempboxa{\algocf@captiontext{#1}{#2}}%
  \ifdim\wd\@tempboxa >\hsize
    \hskip .5\algomargin%
    \parbox[t]{\hsize}{\algocf@captiontext{#1}{#2}}
  \else%
    \global\@minipagefalse%
    \hbox to\hsize{\box\@tempboxa}
  \fi%
  \addtolength{\hsize}{-\algomargin}%
}
\makeatother

\newcommand{\R}{\mathbb{R}}
\newcommand{\p}{\mathbb{P}}
\newcommand{\E}{\mathbb{E}}
\newcommand{\1}{\mathds{1}}
\newcommand{\dd}{\mathrm{d}}

\usepackage{amsmath,amssymb,bm,xcolor}
\usepackage{enumerate,mathrsfs,graphics,caption,subfig,booktabs,verbatim}

\usepackage{dsfont}

\usepackage{array,multirow}
\newcommand{\STAB}[1]{\begin{tabular}{@{}c@{}}#1\end{tabular}}

\begin{document}

\jname{Biometrika}
\jyear{-} 
\jvol{-} 
\jnum{-} 


\markboth{E. Musta and I. Van Keilegom}{Mixture cure model with mismeasured covariates}

\title{A simulation-extrapolation approach for the mixture cure model with mismeasured covariates}

\author{Eni Musta}
\affil{ORSTAT, {KU Leuven}\\ Naamsestraat 69, 3000 Leuven, Belgium
\email{eni.musta@kuleuven.be}}

\author{Ingrid Van Keilegom}
\affil{ORSTAT, {KU Leuven}\\ Naamsestraat 69, 3000 Leuven, Belgium \email{ingrid.vankeilegom@kuleuven.be}}
\maketitle

\begin{abstract}
We consider survival data from a population with cured subjects in the presence of mismeasured covariates. 
We use the mixture cure model to account for the individuals that will never experience the event and at the same time distinguish between the effect of the covariates on the cure probabilities and on  survival times.  In particular, for practical applications, it seems of interest to assume a logistic form of the incidence and a Cox proportional hazards model for the latency. To correct the estimators for the bias  introduced by the measurement error, we use the simex algorithm, which is a very general simulation based method. It essentially estimates this bias by introducing additional error to the data and then recovers bias corrected estimators through an extrapolation approach. The estimators are shown to be consistent and asymptotically normally distributed {when the true extrapolation function is known}. We investigate their finite sample performance through a simulation study and  apply  the proposed method to analyse the effect of the  prostate specific antigen (PSA) on patients with prostate cancer.
\end{abstract}


\section{Introduction}	
Classical survival analysis methods are designed to deal with time-to-event data in the presence of censoring and covariates.  However, they often fail to address various challenges presented by real-life problems. In recent times, significant advances have been made in adapting and extending traditional methods for handling data with more complex features. In this article we account simultaneously for a cure fraction of the population, referring to those subjects that are immune to the event of interest, and covariates measured with error. Such situations arise frequently in practice. For instance, in cancer studies it is known that some of the patients will never experience recurrence or cancer related death and certain biomarker expressions such as the hemoglobin level or tumor size  cannot be measured precisely. The systolic blood pressure is also known to be an error-prone predictor for the development of the coronary heart disease. Examples of variables that cannot be measured precisely end events that are not experienced by the whole population can also be found in economic and social studies. 
Ignoring both these characteristics in the statistical procedures would most probably lead to incorrect inferences. 

Cure rate models
 were first introduced by \cite{boag49} and \cite{berkson52}, but only quite recently they have attracted attention in the statistical literature and applications. The proposed {models} can be divided in two main categories: mixture cure models and promotion time models (see \cite{AK2018} for a detailed review). The first ones assume that the population consists of two subpopulations, the cured and the susceptible ones, and model separately the incidence (the  probability of being noncured) and the latency (the survival of the noncured subjects) using parametric or nonparametric models. The latter ones have a proportional hazards structure and extend the classical Cox regression model to allow for the survival function to flatten at a level greater than zero. There is no clear indication of which approach is more appropriate but in general, mixture cure models are preferred when one wants to distinguish between variables that affect the cure probability and the survival of the uncured subjects.

On the other hand, there is a vast literature about bias correction methods mainly in regression models with covariates contaminated by measurement error (see \cite{carroll2006measurement}). The classical additive error model is generally accepted and the most common methods to deal with it are the so called functional ones, which do not make any assumption on the distribution of the unobserved true covariates. They
can be divided in three large classes of models: regression calibration, score functional methods and simulation-extrapolation (simex). The latter one is in particular quite appealing because it is a simulation based method and it can be easily adapted to any kind of model. It only requires an estimation method in the absence of measurement error  and can be easily implemented (though computationally more intensive). In survival analysis  it has been applied to the semiparametric  Cox model (\cite{carroll2006measurement}), the marginal hazards model for multivariate failure time data (\cite{greene2004measurement}), the frailty model for clustered survival data (\cite{li2003functional}) etc. 

However, there are only limited studies on cure rate models with measurement error. 
This problem was first addressed by \cite{mizoi2007cure} and \cite{ma2008cure}, who propose a corrected score approach for the parametric and semiparametric promotion time models respectively. Afterwards, the simex procedure was introduced as an alternative estimation method in a more general version of promotion time models  by \cite{bertrand2017inference} and an extensive simulation study was done by \cite{bertrand2017robustness} to compare it with the corrected score approach and get a better understanding on the robustness of the method. {In the context of mixture cure models, the simex algorithm has only been proposed for left-truncated right-censored data when  a transformation model is assumed for the latency (\cite{chen2019semiparametric}). However, \cite{chen2019semiparametric} considers only the case in which the mismeasured covariate affects only the latency and theory is developed for one specific estimation method based on martingale integral representations. In particular, the most commonly used logistic/Cox mixture cure model for right-censored data and the maximum likelihood estimation method (based on the EM algorithm) have not been investigated in presence of measurement error. The popularity of this model} motivates us to search for solutions to correct  estimates for the biases induced by the measurement error.

Here we propose a simex approach for a general mixture cure model with a parametric form of the incidence and a semiparametric model for the latency. {Any estimation method in the absence of measurement error can be used within the simex algorithm.} We focus mainly on the logistic/Cox setting, given its practical relevance,  but the proposed procedure and the asymptotic theory hold for other  mixture cure models as well,  provided that the considered estimation method in the absence of measurement error satisfies certain conditions. In particular, these conditions are satisfied for the maximum likelihood 
estimator introduced in \cite{ST2000} and the presmoothing approach proposed by \cite{MPK20}. We use both these estimators in the simex procedure  and compare them through a simulation study. In contrast to the previously considered promotion time models, here we find that if the mismeasured covariate affects only one of the two components (incidence or latency), the estimation of the other component remains undisturbed even if the variables are correlated.  However, the use of the simex algorithm to correct for the bias, not always leads to better results in terms of mean squared error. The decision on whether to choose simex over the naive approach (ignoring the bias) depends on a number of factors. In particular, a large  sample size, a strong effect of the covariate, a relatively large measurement error and low censoring favour the use of the simex approach.

The article is organized as follows. We start by describing a general parametric/semiparametric mixture cure model with measurement error in Section \ref{sec:model} and then  explain the simex estimation procedure in Section \ref{sec:method}. Asymptotic properties of the estimators are presented in Section \ref{sec:asymptotics}, while their practical performance for the logistic/Cox mixture cure model is demonstrated through simulation studies in Section \ref{sec:simulations}. Finally, in Section \ref{sec:application},  we apply the proposed 
method to a prostate cancer dataset to account for measurement error in the values of the prostate specific antigen.

\section{Mixture cure model with measurement error}
\label{sec:model}
Suppose we are interested in the time $T$ until a certain event happens. In contrast to classical survival analysis, in cure models it is possible to have $T=\infty$ (the event never happens), reflecting the presence of a {cure} fraction.  On the other hand, a finite survival time corresponds to susceptible subjects that will experience the event at some time point. If we indicate by $B$ the uncured status, i.e. $B=\1_{\{T<\infty\}}$, then we can write
\[
T=BT_0+(1-B)\infty,
\]
with $T_0$ representing the survival time for an uncured individual. The challenge of dealing with this type of models arises from the fact that, because of finite censoring times, it is impossible to completely separate the two groups. To be precise, if $C$ denotes the censoring time, then we only observe the follow-up time $Y=\min(T,C)$ and the censoring indicator $\Delta=\1_{\{T\leq C\}}$. Hence, for  the observations with $\Delta=0$, we do not know whether they are cured or susceptible. 
In addition to the {cure} fraction and censoring, it is desirable to also account for the impact of certain covariates on the time to event variable. Let $(X^T,Z^T)^T$ a $(p+q)$-dimensional vector of  covariates, where $x^T$ denotes the transpose of the vector $x$. The advantage of mixture cure models with respect to promotion time models is that they can distinguish between the covariates $X$, which affect the cure rate, and $Z$, which affect the survival of the uncured subjects, i.e. 
\begin{equation*}
\label{eqn:condition_X_Z}
\p(T=\infty|X,Z)=\p(T=\infty|X)\qquad \text{ and } \qquad\p(T<\infty|X,Z)=\p(T<\infty|Z).
\end{equation*}
However, it is possible for $X$ and $Z$ to be the same or share some of the components. As commonly done in studies of cure models, we assume that  the censoring time and the survival time are independent given the covariates 
\begin{equation}
\label{eqn:CI1}
T\perp C| (X,Z),
\end{equation}
which is equivalent to requiring  $T_0\perp (C,X) | Z$ and $B\perp (C,T_0,Z)|X$ (see Lemma 1 in Appendix A of \cite{MPK20}). 

In this paper we deal with situations in which some of the continuous covariates included in $X$ and/or $Z$ are subject to measurement error. For ease of notation and interpretation  we define the vector of unique covariates $(E^{{(1)}^T},E^{{(2)}^T},E^{{(3)}^T})^T\in\R^{p+q_1}$ where $E^{(1)}$ denotes the covariates in $X$ that are not present in $Z$, $E^{(2)}$ denotes the common components of $X$ and $Z$, $E^{(3)}$ denotes the covariates in $Z$ that are not present in $X$ {and $q_1$ is the number of covariates in  $E^{(3)}$}. In other words, we are removing the repeated covariates from the vector $(X^T,Z^T)^T$ without loosing any information. In the presence of measurement error, instead of $(E^{{(1)}^T},E^{{(2)}^T},E^{{(3)}^T})^T$, we observe $W=(W^{{(1)}^T},W^{{(2)}^T},W^{{(3)}^T})^T$ such that 
\begin{equation}
\label{eqn:error}
W=\left(E^{{(1)}^T},E^{{(2)}^T},E^{{(3)}^T}\right)^T+U
\end{equation}
where $U\in\R^{p+q_1}$ is the vector of measurement errors. We assume that $U$ is independent of {$(X,Z,T,C)$} and it follows a continuous distribution with mean zero and known variance matrix $V$. The elements of $V$  corresponding to covariates with no measurement error (including non-continuous covariates) are set to zero. However, no parametric assumption is made on the distribution of the errors. {In particular, the measurement error is not required to be normally distributed.} 

We consider a general mixture cure model with a parametric form of the incidence and a semiparametric model for the latency. To be precise,  the cure probability of a subject with covariate $x$ is 
$$
\pi_0(x)=1-\phi(\gamma_0,x)
$$
for some {known} function  $\phi:\,\R^{p}\times\R^p\mapsto [0,1]$ and $\gamma_0\in\R^p$, while the conditional survival function of the noncured subjects $S_u(\cdot|z)$ depends on a parametric component $\beta_0$  and a nonparametric non-decreasing function $\Lambda_0$ (for example the cumulative baseline hazard). {As a result, the conditional survival function corresponding to $T$  is then
$$
S(t|x,z)=\p(T>t|{X=x, Z=z})=1-\phi(\gamma_0,x)+\phi(\gamma_0,x)S_u(t|z).
$$} The logistic model, where
\begin{equation}
\label{eqn:logistic}
\phi(\gamma,x)=\frac{e^{\gamma^Tx}}{1+e^{\gamma^Tx}},
\end{equation}
is perhaps the most common one when a parametric form of the cure probability is adequate. On the other hand, the Cox proportional hazards model {(\cite{Cox72})}
\begin{equation}
\label{eqn:cox}
S_u(t|z)=\exp\left\{-\Lambda_0(t)\exp(\beta^T_0z)\right\},
\end{equation}
and the accelerated failure time model 
\[
S_u(t|z)=\exp\left\{-\Lambda_0\left(\exp\left(\beta^T_0z\right)t\right)\right\},
\]
where $\Lambda_0$ is the baseline cumulative hazard function, are both widely used semiparametric modelling approaches for the latency. 
However, our methodology applies  more in general to parametric/semiparametric mixture cure models provided that an estimation method for the case without measurement error is available. 
The goal is to estimate {the true parameters} $\gamma_0$, $\beta_0$ and $\Lambda_0$ on the basis of  
$n$ i.i.d. observations $(Y_1,\Delta_1,W_1),\dots,(Y_n,\Delta_n,W_n)$, knowing the variance matrix $V$ of the measurement error. In the next section we propose a simulation-extrapolation approach designed to reduce the bias due to the measurement error.
\section{Methodology}
\label{sec:method}
The basic idea behind the simex algorithm is that we can gain insights on how the measurement error affects the estimators by creating artificial data with increasing levels of measurement error and  estimating the parameters as if there was no error. The obtained information is then used in the second step to recover the bias corrected estimators through an extrapolation approach. Next we describe the details of this procedure. 

\bigskip
\textit{Step 1. (Simulation)} We choose $K$ levels of added noise $\lambda_1,\dots,\lambda_K\geq 0$ and for each of them we generate a large number $B$ of artificially contaminated samples. To be precise, for each $\lambda\in \{\lambda_1,\dots,\lambda_K\}$ and $b\in\{1,\dots,B\}$, we simulate independent identically distributed variables $\{\tilde{U}_{b,i}\}_{i=1}^n$,  independently of the observed data and with distribution $ N_D(0,I_D)$, where $D=p+q_1$ is the dimension of the vector $W$. Afterwards, we construct new covariates 
\[
W_{i,\lambda,b}=W_i+(\lambda V)^{1/2}\tilde{U}_{i,b},
\] where $V$ is the covariance matrix of the error in \eqref{eqn:error}. Distributions different from Gaussian can be used too but here we focus on normal errors. {The mixture model satisfied by the new covariates $W_{i,\lambda,b}$ 
\[
\begin{split}
&S\left(t\,\bigg|\left(W_{i,\lambda,b}^{(1)},W_{i,\lambda,b}^{(2)}\right),\left(W_{i,\lambda,b}^{(2)},W_{i,\lambda,b}^{(3)}\right)\right)\\
&=1-\phi\left(\gamma_\lambda,\left(W_{i,\lambda,b}^{(1)},W_{i,\lambda,b}^{(2)}\right)\right)+\phi\left(\gamma_\lambda,\left(W_{i,\lambda,b}^{(1)},W_{i,\lambda,b}^{(2)}\right)\right)S_{u,\lambda}\left(t\,\bigg|\left(W_{i,\lambda,b}^{(2)},W_{i,\lambda,b}^{(3)}\right)\right),
\end{split}
\]
is characterized by the parameters  $\gamma_\lambda$, $\beta_\lambda$ and $\Lambda_\lambda$.
Using $\{Y_i,\Delta_i,W_{i,\lambda,b}\}_{i=1}^n$ we estimate $\gamma_\lambda$, $\beta_\lambda$ and $\Lambda_\lambda$, }
 as if there was no measurement error,  obtaining $\hat{\gamma}_{\lambda,b}$, $\hat{\beta}_{\lambda,b}$ and $\hat{\Lambda}_{\lambda,b}$. The latter one is an estimator of $\Lambda_0$ over some compact interval $[0,\tau]$. Any available estimation method can be used. For example, in the logistic/Cox mixture cure model, the maximum likelihood estimation (\cite{ST2000,cai_smcure}) or the presmoothing approach proposed by \cite{MPK20} can be considered. 

At the end, for each level of contamination, the average values of all the $B$ estimates are calculated:
\begin{equation}
\label{eqn:averages}
\hat\gamma_\lambda=\frac{1}{B}\sum_{b=1}^B\hat{\gamma}_{\lambda,b},\qquad \hat\beta_\lambda=\frac{1}{B}\sum_{b=1}^B\hat{\beta}_{\lambda,b}\qquad\text{ and }\qquad \hat\Lambda_\lambda(t)=\frac{1}{B}\sum_{b=1}^B\hat{\Lambda}_{\lambda,b}(t).
\end{equation}
Note that, if the estimators $\hat{\Lambda}_{\lambda,b}$ are piecewise constant  with jumps at the observed event times, then also $\hat{\Lambda}_{\lambda}$ is piecewise constant  with jumps at the observed event times.
The parameters to be chosen in this step are $K$, the $\lambda$s and $B$. Common values are $K=5$, $\lambda\in\{0,0.5,1,1.5,2\}$ and $B=50$ (\cite{carroll1996asymptotics,cook1994simulation}).

\bigskip
\textit{Step 2. (Extrapolation)} Note that, by independence, the covariance matrix of the simulated covariates $W_{i,\lambda,b}$ is 
\[
\mathrm{var}(W_{i,\lambda,b}|X_i)=\mathrm{var}(W_i|X_i)+\lambda V=(1+\lambda)V.
\]
This means that the variance has been inflated by a factor $1+\lambda$ and that the ideal case of no measurement error corresponds to $\lambda=-1$ (adding `negative' variance). Hence, the idea is to model the relationship between $\lambda$ and the estimators $\hat\gamma_\lambda$, $\hat{\beta}_\lambda$, $\hat{\Lambda}_\lambda$ by fitting a regression function and then extrapolate to $\lambda=-1$. First, an extrapolant function needs to be chosen (e.g. linear, quadratic or fractional) for each component of $\hat\gamma_\lambda$, $\hat\beta_\lambda$, $\hat{\Lambda}_\lambda$ as a function of $\lambda$. For example, for the quadratic case and $\lambda\in\{\lambda_1,\dots,\lambda_K\}$, we have
\[
\begin{split}
\hat{\gamma}_{\lambda,j}=g_{\gamma,j}(a_{\gamma_j}^*,\lambda)+\epsilon_{\gamma,\lambda,j}&=a_{\gamma_j,1}^*+a_{\gamma_j,2}^*\lambda+a_{\gamma_j,3}^*\lambda^2+\epsilon_{\gamma,\lambda,j},\qquad j=1,\dots,p\\
\hat{\beta}_{\lambda,j}=g_{\beta,j}(a_{\beta_j}^*,\lambda)+\epsilon_{\beta,\lambda,j}&=a_{\beta_j,1}^*+a_{\beta_j,2}^*\lambda+a_{\beta_j,3}^*\lambda^2+\epsilon_{\beta,\lambda,j},\qquad j=1,\dots,q\\
\hat{\Lambda}_{\lambda}(t)=g_{\Lambda,t}(a_{t}^*,\lambda)+\epsilon_{\Lambda,\lambda,t}&=a_{t,1}^*+a_{t,2}^*\lambda+a_{t,3}^*\lambda^2+\epsilon_{\Lambda,\lambda,t},\qquad\,\,\,\quad t\in[0,\tau],
\end{split}
\]
where $\epsilon_{\beta,\lambda,j}$, $\epsilon_{\gamma,\lambda,j}$ and $\epsilon_{\Lambda,\lambda,t}$ are the error terms in the extrapolant model, assumed to have mean zero and to be independent.  Estimators  $\hat{a}_{\gamma_j}=(\hat{a}_{\gamma_j,1},\hat{a}_{\gamma_j,2},\hat{a}_{\gamma_j,3})$, 
$\hat{a}_{\beta_j}=(\hat{a}_{\beta_j,1},\hat{a}_{\beta_j,2},\hat{a}_{\beta_j,3})$ and $\hat{a}_{t}=(\hat{a}_{t,1},\hat{a}_{t,2},\hat{a}_{t,3})$ of the unknown parameters of the extrapolant function are obtained by fitting the previous regression models using the method of  least squares. Finally, the simex estimators are defined by
\[
\begin{split}
&\hat\gamma_{j,\mathrm{simex}}=\lim_{\lambda\to -1} g_{\gamma,j}(\hat{a}_{\gamma_j},\lambda),\qquad j=1,\dots,p,\\
&\hat\beta_{j,\mathrm{simex}}=\lim_{\lambda\to -1} g_{\beta,j}(\hat{a}_{\beta_j},\lambda),\qquad j=1,\dots,q,\\
&\hat{\Lambda}_{\mathrm{simex}}(t)=\lim_{\lambda\to -1} g_{\Lambda,t}(\hat{a}_{t},\lambda),\qquad t\in[0,\tau].
\end{split}
\]
If the initial estimators $\hat{\Lambda}_{\lambda,b}$ are piecewise constant  with jumps at the observed event times, then the extrapolation procedure needs to be applied only for the observed event times $t\in\{T_{(1)},\dots,T_{(m)}\}$. {Equivalently, the procedure can be applied to the jump sizes for  different coefficients $a^*$ and a possibly different extrapolation function (if it is not polynomial). Even though this does not guarantee that the resulting estimator $\hat\Lambda_{\mathrm{simex}}$ is non-decreasing, in practice this is often the case. If one is interested in estimation of $\Lambda_0$ on the whole support and $\hat\Lambda_{\mathrm{simex}}$ is not monotone, an isotonized version of it, using for example the pool-adjacent-violators algorithm (\cite{robertson}),  would be a more reasonable estimate.  However, here we focus on estimation of the parameters $\gamma$, $\beta$ and do not further exploit this aspect.}
Note also that different extrapolation functions lead to different results. Hence it is important to have a good approximation of the true extrapolation function. 
\section{Asymptotic properties}
\label{sec:asymptotics}
\subsection{General results}
In this section we establish some theoretical results regarding the large-sample properties of the proposed  estimators. A drawback of the simex approach is that  consistency and asymptotic normality of the estimators hold only if we knew the true extrapolation function, which is usually not the case in practice. 
 When the true extrapolant function is not known, but an approximation of it is used, the results of Theorems \ref{theo:consistency} and \ref{theo:normality} hold with $\gamma_0$, $\beta_0$, $\Lambda_0(t)$ replaced by $\lim_{\lambda\to-1}g_\gamma(a_\gamma,\lambda)$,  $\lim_{\lambda\to-1}g_\beta(a_\beta,\lambda)$ and  $\lim_{\lambda\to-1}g_\Lambda(a_t,\lambda)$ respectively. {Here $g_\gamma(a_\gamma,\lambda)$ denotes the vector $(g_{\gamma,1}(a_{\gamma_1},\lambda),\dots,g_{\gamma,p}(a_{\gamma_p},\lambda))^T $ and  $g_\beta(a_\beta,\lambda)$, $g_\Lambda(a_t,\lambda)$ are defined similarly.}
We first establish the asymptotic results in a general mixture cure model as described in Section~\ref{sec:model}, assuming that the used estimation method for obtaining $\hat\gamma_{\lambda,b}$, $\hat\beta_{\lambda,b}$, $\hat\Lambda_{\lambda,b}$ (ignoring the measurement error) satisfies certain conditions. Afterwards, we will focus on two  estimation methods for the logistic/Cox mixture cure model and show that the required conditions are met. All the proofs can be found in the Supplementary Material.

For a fixed $\lambda>0$ consider observations $(Y,\Delta,W_\lambda)$, where $W_\lambda=W+(\lambda V)^{1/2}\tilde{U}$ and the mixture cure model  with conditional survival
\[
S(t|W_\lambda)=1-\phi\left(\gamma_\lambda,\left(W_\lambda^{(1)},W_\lambda^{(2)}\right)\right)+\phi\left(\gamma_\lambda,\left(W_\lambda^{(1)},W_\lambda^{(2)}\right)\right)S_{u,\lambda}\left(t\,\bigg|\left(W_\lambda^{(2)},W_\lambda^{(3)}\right)\right),
\]
where, as mentioned in Section \ref{sec:model}, the decomposition of $W_\lambda$ in three components corresponds to the covariates that influence only the cure probability, those that are common for the incidence and the latency and the ones that affect only the latency. The  survival  of the uncured subject $S_{u,\lambda}$ depends on the regression parameters $\beta_\lambda$ and the nonparametric function $\Lambda_\lambda$. Suppose we have an estimation method that provides estimates $\hat\gamma_\lambda$, $\hat\beta_\lambda$ and $\hat\Lambda_\lambda$, the latter one being a non-decreasing function.  The following conditions will be needed in order to establish the asymptotic results. 
\begin{itemize}
	\item[(A1)] With probability one and for some $\tau>0$ we have
	\[
	\Vert\hat\gamma_\lambda-\gamma_\lambda\Vert\to 0,\qquad \Vert\hat\beta_\lambda-\beta_\lambda\Vert\to 0\qquad\text{ and} \qquad\sup_{t\in[0,\tau]}|\hat\Lambda_\lambda(t)-\Lambda_\lambda(t)|\to 0
	\]
	as $n\to\infty$, i.e. the estimators are strongly consistent. By $\Vert\cdot\Vert$ we denote the Euclidean norm.
	\item[(A2)] For $\mathfrak{m}<\infty$, define
	\[
	\mathcal{H}_{\mathfrak{m}}=\left\{h=(h_1,h_2,h_3)\in BV[0,\tau]\times\R^p\times\R^q\,:\,\Vert h\Vert_H= \Vert h_1\Vert_v+\Vert h_2\Vert+\Vert h_3\Vert\leq \mathfrak{m} \right\}
	\]
	where {$BV[0,\tau]$ denotes the space of functions of bounded variation on $[0,\tau]$, }$\Vert h_1\Vert_v=|h_1(0)|+V_0^\tau(h_1)$ and $V_0^\tau(h_1)$ denotes the total variation of $h_1$ over $[0,\tau]$. Uniformly over $h\in\mathcal{H}_{\mathfrak{m}}$ we have
	\[
	\begin{split}
	&h^T_2(\hat{\gamma}_\lambda-\gamma_\lambda)+h_3^T(\hat{\beta}_\lambda-\beta_\lambda)+\int_0^\tau h_1(s)\dd(\hat\Lambda_\lambda-\Lambda_\lambda)(s)\\
	&=\frac{1}{n}\sum_{i=1}^n\Psi_\lambda(Y_i,\Delta_i,W_{i,\lambda},h_1,h_2,h_3)+o_P(n^{-1/2})
	\end{split}
	\]
	for some function $\Psi_\lambda$ such that $\E[\Psi_\lambda(Y,\Delta,W_{\lambda},h_1,h_2,h_3)]=0$ and for fixed $\lambda$, the class
	\[
	\left\{(y,\delta,w)\mapsto \Psi_\lambda(y,\delta,w,h_1,h_2,h_3)\,:\,(h_1,h_2,h_3)\in\mathcal{H}_{\mathfrak{m}}\right\}
	\]
	is uniformly bounded and  Donsker.
\end{itemize}
Moreover, in what follows,  we assume that the extrapolant functions $g(a,\lambda)$ are such that the matrix $\dot{g}(a,\lambda)$ of partial derivatives with respect to the elements of $a$ is bounded and continuous {at the true parameters $a^*$} and has full rank, i.e. $\dot{g}({a^*},\lambda)^T\dot{g}({a^*},\lambda)$ is invertible.

\begin{theorem}
	\label{theo:consistency}
	Suppose that condition (A1) is satisfied and that $\Lambda_0$ is continuous. If the measurement error variance and the true extrapolant functions are known then, with probability one,
	\[
	\Vert\hat\gamma_{\mathrm{simex}}-\gamma_0\Vert\to 0,\qquad \Vert\hat\beta_{\mathrm{simex}}-\beta_0\Vert\to 0\qquad\text{ and} \qquad\sup_{t\in[0,\tau]}|\hat\Lambda_{\mathrm{simex}}(t)-\Lambda_0(t)|\to 0.
	\]
\end{theorem}

\begin{theorem}
	\label{theo:normality}
	Suppose that conditions (A1)-(A2) are satisfied and that $\Lambda_0$ is continuous. If the measurement error variance and the true extrapolant functions are known, then $n^{1/2}(\hat\gamma_{\mathrm{simex}}-\gamma_0)$ converges in distribution to $N(0,\Sigma_\gamma)$ and $n^{1/2}(\hat\beta_{\mathrm{simex}}-\beta_0)$ converges in distribution to $N(0,\Sigma_\beta)$, with $\Sigma_\gamma$ and $\Sigma_\beta$ as in (S2) and (S3) {in the Supplementary Material}. Moreover, $n^{1/2}(\hat\Lambda_{\mathrm{simex}}-\Lambda_0)$ converges weakly {in $l^\infty([0,\tau])$} to a mean zero Gaussian process $\mathcal{G}$ defined in (S4). 
\end{theorem}
The proofs of Theorems~\ref{theo:consistency} and \ref{theo:normality} follow the usual arguments for simex estimators. In particular, consistency relies mainly on the consistency of the  estimators  for each $\lambda$ and consistency of the estimated extrapolant functions. Moreover, the i.i.d. representation in condition (A2) and the expressions in \eqref{eqn:averages} allow us to obtain convergence to a Gaussian process for any $\lambda$. Finally, the asymptotic normality of the simex estimators follows by the delta method. Details of the proofs can be found in the Supplementary Material.

\subsection{Example: logistic/Cox mixture cure model}
\label{sec:Cox}
The logistic/Cox mixture cure model is perhaps the most commonly used one for studying survival data in the presence of a cure fraction. It assumes that the function $\phi(\gamma,x)$ is as in~\eqref{eqn:logistic}, 
where the first component of $x$ is equal to one and $\gamma_1$ corresponds to the intercept.
On the other hand, the distribution of the uncured subjects  follows a Cox proportional hazards model
as in~\eqref{eqn:cox}, where $\Lambda_0$ is the baseline cumulative hazard,  $\beta^T_0Z$ does not contain an intercept and the matrix $\mathrm{var}(Z)$ is assumed to have full rank for the Cox model to be identifiable. The classical estimator in this setting is the maximum likelihood estimator proposed by \cite{ST2000} and implemented in the $\mathrm{R}$ package \texttt{smcure}. The estimator is computed through the expectation maximization algorithm because of the unobserved latent variable $B$ and its asymptotic properties  are investigated in \cite{Lu2008}. 
Recently, an alternative estimation procedure relying on  presmoothing  was proposed by \cite{MPK20}. It uses  a preliminary nonparametric estimator for the cure probabilities and ignores the Cox model when estimating $\gamma_0$. It is shown through simulations that, if the interest is focused on estimation of the parameters of the incidence, this method usually performs better that the maximum likelihood estimator. However, both methods lead to very similar results when estimating the latency. 
Next we show that these two estimators satisfy our conditions (A1)-(A2) and as a result, both procedures can be used in the SIMEX algorithm leading to consistent and square-root  convergent estimators.  
\begin{theorem}
	\label{theo:smcure}
	Consider the maximum likelihood estimation method proposed by \cite{ST2000}.
	Assume that conditions 1-4  in \cite{Lu2008} are satisfied. Then our conditions (A1)-(A2) above hold with $\Psi_\lambda(y,\delta,w,h_1,h_2,h_3)$ as in (S9) {in the Supplementary Material}.
\end{theorem}
\begin{theorem}
	\label{theo:bb}
	Consider the estimation method proposed by \cite{MPK20} and assume that their assumptions  (C1)-(C4), {(AC2), (AC5)-(AC7)} are satisfied. Then our conditions (A1)-(A2) above hold with $\Psi_\lambda(y,\delta,w,h_1,h_2,h_3)$ as in (S11) {in the Supplementary Material}.
\end{theorem}
In order for the mixture cure model to be identifiable,  $T_0$ should have compact support $[0,\tau_0]$ such that $\inf_{x,z}\p(C>\tau_0|X=x,Z=z)>0$.  Hence, $\tau$ in our conditions $(A1)-(A2)$ is equal to $\tau_0$. In practice cure rate models are used when there is a long follow-up beyond the largest observed event time $T_{(m)}$ and 
the zero-tail constraint is applied, i.e. the censored subjects with follow-up time larger than  $T_{(m)}$ are considered cured. For being able to develop the asymptotic theory, in \cite{Lu2008}  it is assumed that $
\inf_z\p(T_0=\tau_0|Z=z)>0,
$
while \cite{MPK20} argue that  this assumption can be avoided thanks to the presmoothing step.

\section{Numerical study}
\label{sec:simulations}
\subsection{Setup}
In this section we investigate the finite-sample behaviour of the simex method in the logistic/Cox mixture cure model. The two estimation approaches considered in Section~\ref{sec:Cox} are used within the simex algorithm and compared with each other in the context of mismeasured covariates.  
Results for a variety of models and scenarios are presented in the next subsections. We try to cover a wide range of situations and capture the effect of the cure rate, censoring rate, sample size and measurement error variance. Unless stated otherwise, the error distribution is Gaussian and the used extrapolation function is quadratic, which seems to be a good compromise in terms of bias and variance (\cite{cook1994simulation,carroll2006measurement,li2003functional,bertrand2017inference}). Finally, we also briefly investigate the robustness of the method with respect to the extrapolation function, misspecification of the error distribution and variance. In all the simulation studies, for the simex method, we choose $B=50$, $K=5$, $\lambda\in\{0,0.5,1,1.5,2\}$ (as these seem to be quite common choices in the literature) and for each setting $500$ simulated datasets were used to compute the bias, variance and mean squared error (MSE) of the estimators. We also compare the bias corrected estimators with the naive estimators, which do not take the measurement error into account. {The bandwidth for the estimator based on presmoothing is chosen as in \cite{MPK20}, i.e.  the cross-validation optimal bandwidth
	for estimation of the conditional distribution $H(t|x)$ for $t\leq  Y_{(m)}$ truncated from above at $2$, where $Y_{(m)}$ is the largest uncensored observation and $x$ is the continuous covariate affecting the incidence (standardized). To reduce  computational time, we compute this bandwidth only once for the initial dataset and use the same for the data with added noise. We observed that not updating the bandwidth for each $b\in\{1,\dots,B\}$ and $\lambda\in\{0.5,1,1.5,2\}$ does not have a significant impact on the final results. } Moreover, we assume to know the standard deviation of the error, which is usually not the case in practice. In such situations, a preliminary step of variance estimation is required before applying the simex procedure (see for example \cite{bertrand2019flexible}).
\subsection{One mismeasured covariate}
We start by considering a simplified model in which there is only one covariate of interest, measured with error, affecting both the cure probability and the survival of the uncured subjects. 

\textit{Model 1.} Both incidence and latency depend on one covariate $X$, which is a standard normal random variable. We generate the cure status $B$ as a Bernoulli random variable with success probability $\phi(\gamma,x)=1/(1+\exp(-\gamma_1-\gamma_2x))$. The survival times for the uncured observations are generated according to a Weibull proportional hazards model
\[
S_u(t|x)=\exp\left(-\mu t^\rho\exp(\beta^Tx)\right),
\]
and are truncated at $\tau_0=7$ for $\rho=1.75$, $\mu=1.5$ and $\beta=1$. The censoring times are independent from $X$ and $T$. They are generated from the exponential distribution with parameter $\lambda_C$ and are truncated at $\tau=9$. Various choices  of the parameters $\gamma$ and $\lambda_C$ with the corresponding cure and censoring rates can be found in Table \ref{tab:model_1}.  Here and in what follows, the truncation of the  survival times and censoring times is done in such a way that  $\tau_0<\tau$ and it is unlikely to observe an event time at $\tau_0$. This mimics real-life situations in which cure models are adequate.
$X$ is measured with error, i.e. instead of $X$ we observe $W=X+U$, where $U\sim N(0,v^2)$.
\begin{table}[h]
	\centering
	\scalebox{0.85}{
		\begin{tabular}{ccccccccc}
			& & & & & & &&\\[-8pt]
			Setting & $\gamma_2$ & Scenario & Cure rate & $\gamma_1$ & Cens. rate  & $\lambda_C$ & Cens. level& Plateau\\[2pt]
			& & & & & & & &\\[-8pt]
			1 & $0.1$  &1 &$20\%$ &$1.4$  & $1$  & $0.09 $ & $25\%$&$14\%$\\
			&   & & &  & $2$  & $0.3 $ & $35\%$&$7\%$\\
			& & & & & & & &\\[-8pt]
			& &2 &$50\%$ &$0$  & $1$  & $0.13 $ & $55\%$&$32\%$\\
			&  & & &  & $2$  & $0.5 $ & $65\%$ &$15\%$\\
			& & & & & & & &\\[-8pt]
			2 & $0.5$  &1 &$20\%$ &$1.4$  & $1$  & $0.07 $ & $25\%$&$16\%$\\
			&   & & &  & $2$  & $0.26 $ & $35\%$&$9\%$\\
			& & & & & & & &\\[-8pt]
			& &2 &$50\%$ &$0$  & $1$  & $0.15 $ & $55\%$&$31\%$\\
			&  & & &  & $2$  & $0.6 $ & $65\%$&$14\%$\\
			& & & & & & & &\\[-8pt]
			3 & $2$  &1 &$20\%$ &$2.2$  & $1$  & $0.1 $ & $25\%$&$15\%$\\
			&   & & &  & $2$  & $0.33 $ & $35\%$ &$9\%$\\
			& & & & & & & & \\[-8pt]
			& &2 &$50\%$ &$0$  & $1$  & $0.2 $ & $55\%$ &$33\%$\\
			&  & & &  & $2$  & $0.7 $ & $65\%$&$16\%$\\
	\end{tabular}}
	\caption{Parameter values and characteristics of each scenario for Model 1.}
\label{tab:model_1}
\end{table}

Results for sample size $n=200$ ($n=400$) and measurement error variance $v^2=0.7^2$ are given in Table \ref{tab:results1_200} (Table~\ref{tab:results1_400} in the Supplementary Material). This corresponds to a large error situation since the ratio between the standard deviation of the error and the standard deviation of the covariate is $0.7$. Below we will consider also settings with smaller measurement error.

First of all, we observe that in the presence of measurement error there is usually no advantage of using the presmoothing approach instead of maximum likelihood estimation. In particular, when the bias induced by the measurement error is large, it seems that the estimator based on presmoothing is more affected for both the naive and the simex method. Moreover, most of the time the bias is observed only for the coefficients that correspond to the variables measured with error. As expected, in all cases, the simex algorithm reduces this bias at the price of a larger variance. 
In terms of mean squared error, it is better to use the naive approach for coefficients that are small in absolute value (the case of $\gamma_2$ in setting 1), while the simex method is preferred when  the absolute value of the coefficient is large (i.e. the covariate has a greater effect on the cure/survival). In this setting, for $n=200$,  $\gamma_2=0.5$ seems to be a borderline case, meaning that the simex method performs better when the censoring rate is low, while the naive method  has smaller MSE when the censoring rate is high. In addition, results show that when the coefficient of a mismeasured covariate is large, there might be induced bias even for the intercept, which is also corrected by the simex algorithm. As the sample size increases, the bias created by the measurement error increases but the variance decreases for both naive and simex estimators. Furthermore, the  advantage of using simex instead of ignoring the bias becomes more significant. At the same time, the threshold absolute value of a coefficient for which bias correction leads to better MSE decreases (simex is preferred for $\gamma_2=0.5$ in setting 2, which was a borderline case for $n=200$).

\begin{table}[t]
	\centering
	\scalebox{0.85}{
		\begin{tabular}{ccrrrrrrrrrrrr}
			&&\multicolumn{3}{c}{naive - 1}&\multicolumn{3}{c}{naive - 2}&\multicolumn{3}{c}{simex - 1}&\multicolumn{3}{c}{simex - 2}\\
			& Par. &  Bias & Var. & MSE & Bias & Var. & MSE & Bias & Var. & MSE&Bias & Var. & MSE\\[2pt]
			& & & & & & & & & && &&\\[-8pt]
			$1/1/1$  & $\gamma_1 $ & $2.4 $  & $3.8 $ & $ 3.8 $ & $1.5  $ & $3.8 $ & $ 3.8 $ & $2.4 $ & $3.8  $ & $  3.9$ &$1.4$&$4.0$&$4.0$\\
			& $\gamma_2 $ & $-3.5  $  & $2.6  $ & $ 2.7  $ & $-2.7 $  & $ 2.5 $ & $ 2.5$ & $ -1.2 $ &$4.8$&$4.8$&$-0.2$&$5.1$&$5.1$\\
			& $\beta $ & $-43.3  $ & $0.7  $ & $19.4 $ & $-43.3  $ & $ 0.7  $ & $ 19.4 $ & $-18.8 $ & $1.9  $ & $ 5.5$ & $-18.9  $& $1.9  $ &$5.5$\\
			& & & & & & & & & && &&\\[-8pt]
			$1/1/2$  & $\gamma_1 $ & $3.9 $  & $5.5$ & $ 5.6$ & $ 0.5 $ & $ 5.1$  & $ 5.2 $&  $ 3.6  $ & $ 5.6  $ & $ 5.7  $ &$-0.3 $&$6.2 $&$6.2 $\\
			& $\gamma_2 $ & $ -3.0  $  & $ 4.0  $ & $ 4.1  $ & $0.3 $  & $  3.8 $ & $3.9  $ & $ -0.2 $ &$7.3 $&$7.3 $&$ 4.5$&$ 8.6$&$8.8 $\\
			& $\beta $ & $  -42.4 $ & $0.9 $ & $ 18.9 $ & $-42.8   $ & $ 0.9   $ & $19.2 $ & $-18.2  $ & $ 2.4  $ & $5.7  $ & $-18.6   $& $ 2.4 $ &$ 5.9$\\
			& & & & & & & & & && &&\\[-8pt]
			$1/2/1$ & $\gamma_1 $ & $0.7 $  & $2.4$ & $ 2.4$ & $0.2  $ & $2.4 $ &  $  2.4 $ & $ 0.7 $ & $2.4   $ & $ 2.4  $ &$ 0.2$&$2.5 $&$2.5 $\\
			& $\gamma_2 $ & $ -4.0  $  & $ 1.6  $ & $1.8   $ & $ -3.6$  & $  1.4 $ & $ 1.5 $ & $-1.9  $ &$3.0 $&$3.0 $&$-1.3 $&$2.7 $&$ 2.7$\\
			& $\beta $ & $ -42.5  $ & $ 1.2$ & $19.3  $ & $  -42.6 $ & $ 1.2   $ & $19.3 $ & $ -18.1 $ & $ 3.3  $ & $6.6  $ & $-18.2   $& $ 3.3 $ &$ 6.6$\\
			& & & & & & & & & && &&\\[-8pt]
			$1/2/2$  & $\gamma_1 $ & $1.6$  & $4.2$ & $ 4.2$ & $ -0.3 $ & $4.0 $ &  $ 4.0  $ & $ 1.3 $ & $ 4.3  $ & $ 4.3  $ &$-0.2 $&$4.7 $&$4.7 $\\
			& $\gamma_2 $ & $-3.0$ &$2.8   $  & $  2.9 $ & $ -2.1  $ & $2.4 $  & $ 2.5 $ & $ -0.4 $ & $ 5.0 $ &$5.0 $&$0.4 $&$ 5.2$&$ 5.2$\\
			& $\beta $ & $ -41.9  $ & $1.9 $ & $ 19.5 $ & $ -42.3  $ & $  1.8  $ & $19.8 $ & $-18.0  $ & $ 4.9  $ & $ 8.1 $ & $-18.4   $& $ 4.7 $ &$8.1 $\\
			& & & & & & & & & && &&\\[-8pt]
			$2/1/1$  & $\gamma_1 $ & $-0.3 $  & $3.7$ & $ 3.7$ & $ -1.0 $ & $3.6 $ &  $  3.7 $ & $1.4  $ & $   3.9$ & $ 4.0  $ &$0.4 $&$4.1 $&$ 4.1$\\
			& $\gamma_2 $ & $ -15.9  $  & $2.3   $ & $ 4.9  $ & $ -16.6$  & $ 2.2  $ & $ 5.0 $ & $ -3.9 $ &$4.5 $&$4.7 $&$ -5.1$&$4.7 $&$5.0 $\\
			& $\beta $ & $ -44.2  $ & $ 0.7$ & $ 20.2 $ & $ 44.2  $ & $  0.7  $ & $ 20.2$ & $-19.7  $ & $ 2.0  $ & $5.8  $ & $ -19.7  $& $2.0  $ &$ 5.8$\\
			& & & & & & & & & && &&\\[-8pt]
			$2/1/2$  & $\gamma_1 $ & $2.1 $  & $4.8$ & $4.8$ & $ -0.5 $ & $4.7 $ &  $4.7   $ & $ 3.8 $ & $ 5.2  $ & $5.3   $ &$0.8 $&$ 5.8$&$5.8$\\
			& $\gamma_2 $ & $ -16.0  $  & $ 3.6  $ & $ 6.2  $ & $-14.8 $  & $3.4   $ & $ 5.6 $ & $-3.6  $ &$ 7.0$&$ 7.1$&$-1.5$&$7.9 $&$8.0 $\\
			& $\beta $ & $-43.1   $ & $0.9 $ & $ 19.4 $ & $ -43.3  $ & $  0.9  $ & $ 19.6$ & $ -18.6 $ & $  2.4 $ & $  5.9$ & $  -18.9$& $ 2.4 $ &$ 5.9$\\
			& & & & & & & & & && &&\\[-8pt]
			$2/2/1$  & $\gamma_1 $ & $0.8$  & $2.3$ & $ 2.3$ & $0.3  $ & $ 2.3$ &  $ 2.3  $ & $0.8  $ & $ 2.4  $ & $ 2.4  $ &$0.2 $&$2.5 $&$ 2.5$\\
			& $\gamma_2 $ & $ -16.7  $  & $ 1.8  $ & $  4.6 $ & $-17.9 $  & $  1.7 $ & $ 4.9 $ & $ -4.8 $ &$ 3.5$&$ 3.7$&$-6.3 $&$ 3.5$&$ 3.9$\\
			& $\beta $ & $ -43.6  $ & $ 1.3$ & $ 20.3 $ & $ -43.6  $ & $  1.3  $ & $ 20.3$ & $-19.1  $ & $3.5   $ & $ 7.2 $ & $ -19.1  $& $ 3.5 $ &$ 7.2$\\
			& & & & & & & & & && &&\\[-8pt]
			$2/2/2$ & $\gamma_1 $ & $0.6$  & $4.7$ & $ 4.7$  & $-1.7 $ &  $ 4.3  $ & $ 4.3 $ & $ 0.1  $ & $4.8   $ &$4.8 $&$ -2.2$&$4.9 $& $ 4.9 $\\
			& $\gamma_2 $ & $ -15.0  $  & $ 3.4  $ & $ 5.6  $ & $ -16.3$  & $ 3.1  $ & $ 5.8 $ & $ -2.2 $&$ 6.4$&$ 6.5$&$-4.0 $&$6.5 $ &$6.7 $\\
			& $\beta $ & $ -43.1  $ & $2.0 $ & $ 20.6 $ & $ -43.2  $ & $ 1.9   $ & $ 20.5$ & $  -19.3$ & $5.0   $ & $8.8  $ & $ -19.3  $& $ 4.8 $ &$8.5 $\\
			& & & & & & & & & && &&\\[-8pt]
			$3/1/1$  & $\gamma_1 $ & $-33.6 $  & $7.1$ & $ 18.4$ & $-36.6  $ & $ 6.8$ &  $ 20.1  $ & $ -12.1 $ & $13.3   $ & $14.8   $ &$-16.5 $&$12.9 $&$15.6 $\\
			& $\gamma_2 $ & $ -84.6  $  & $6.0   $ & $  77.5 $ & $-88.6 $  & $ 5.4  $ & $ 84.0 $ & $-34.0  $ &$17.0 $&$28.6 $&$-40.3 $&$16.1 $&$ 32.3$\\
			& $\beta $ & $ -48.0  $ & $0.8 $ & $ 23.9 $ & $  -48.0 $ & $  0.8  $ & $23.8 $ & $-23.1  $ & $  2.3 $ & $ 7.6 $ & $ -23.0  $& $ 2.3 $ &$7.6 $\\
			& & & & & & & & & && &&\\[-8pt]
			$3/1/2$  & $\gamma_1 $ & $-31.0$  & $10.6$ & $ 20.3$ & $-37.2  $ & $9.8 $ &  $ 23.7  $ & $-9.0  $ & $ 19.1  $ & $  19.9 $ &$-17.4 $&$19.5 $&$22.6 $\\
			& $\gamma_2 $ & $ -83.7  $  & $ 8.5  $ & $  78.5 $ & $-89.2 $  & $ 7.4  $ & $  86.9$ & $-32.0  $ &$23.6 $&$33.8 $&$ -40.6$&$ 22.4$&$ 38.9$\\
			& $\beta $ & $ -46.8  $ & $1.0 $ & $22.9  $ & $ -46.7  $ & $   1.0 $ & $22.8 $ & $  -21.9$ & $ 2.7  $ & $ 7.4 $ & $ -21.8  $& $ 2.6 $ &$ 7.4$\\
			& & & & & & & & & && &&\\[-8pt]
			$3/2/1$  & $\gamma_1 $ & $0.9 $  & $2.9$ & $2.9$ & $0.4  $ & $2.8 $ &  $ 2.8  $ & $ 1.0 $ & $ 3.8  $ & $3.8   $ &$0.7 $&$ 3.9$&$ 3.9$\\
			& $\gamma_2 $ & $ -88.8  $  & $ 4.0  $ & $  82.9 $ & $-93.6 $  & $  3.8 $ & $91.4  $ & $ -38.6 $ &$11.0 $&$25.9 $&$ -46.1$&$ 11.0$&$32.2 $\\
			& $\beta $ & $ -51.1  $ & $ 1.6$ & $ 27.7 $ & $  -50.9 $ & $ 1.6   $ & $ 27.5$ & $ -26.5 $ & $  4.7 $ & $11.7  $ & $  -26.2 $& $ 4.7 $ &$11.5 $\\
			& & & & & & & & & && &&\\[-8pt]
			$3/2/2$  & $\gamma_1 $ & $1.8 $  & $5.2$ & $ 5.3$ & $-0.8  $ & $4.9 $ &  $ 4.9  $ & $1.8  $ & $  6.9 $ & $  6.9 $ &$-1.0 $&$6.9 $&$6.9 $\\
			& $\gamma_2 $ & $ -87.6  $  & $6.7   $ & $  83.6 $ & $-95.4$  & $ 5.8  $ & $ 96.9 $ & $ -37.0 $ &$19.3 $&$ 33.0$&$ -48.9$&$17.6 $&$41.5 $\\
			& $\beta $ & $ -49.5  $ & $ 2.2$ & $ 26.7 $ & $ -48.9  $ & $ 2.1   $ & $26.0 $ & $ -24.4 $ & $  6.1 $ & $ 12.1 $ & $  -23.7 $& $ 5.7 $ &$ 11.3$\\
	\end{tabular}}
	\caption{Bias, variance and MSE of $\hat\gamma$ and $\hat\beta$ for the naive and simex method  based on the maximum likelihood  (1) or the  presmoothing  (2) approach for Model 1 ($n=200$). The first column gives the setting/scenario/cens. level. All numbers were multiplied by $100$.}
	\label{tab:results1_200}
\end{table}

\subsection{More realistic scenarios}
Through the following four models we try to cover more realistic situations and investigate the effect of the measurement error on the naive and bias corrected estimators.

\textit{Model 2.} Both incidence and latency depend on two independent covariates: $X_1$ has a uniform distribution on the interval $[-1,1]$ and $X_2$ is a Bernoulli random variable with success probability $0.5$. We generate the cure status $B$ as a Bernoulli random variable with success probability  $\phi(\gamma,x)=1/(1+\exp(-\gamma_1-\gamma_2x_1-\gamma_3x_2))$. The survival times for the uncured observations are generated according to a Weibull proportional hazards model
\[
S_u(t|x)=\exp\left(-\mu t^\rho\exp(\beta_1x_1+\beta_2x_2)\right),
\]
and are truncated at $\tau_0$ for $\rho=1.75$ and  $\mu=1.5$. The censoring times are independent from $(X,T)$. They are generated from the exponential distribution with parameter $\lambda_C$ and are truncated at $\tau$.
Instead of $X_1$ we observe $W=X_1+U$, where $U\sim N(0,v^2)$. We consider $v\in\{0.2, 0.4\}$ corresponding to  {small and large} error settings respectively.

\textit{Model 3.} For the incidence we consider two independent covariates: $X_1$ has a uniform distribution on the interval $[-1,1]$ and $X_2$ is a Bernoulli random variable with success probability $0.5$.  The latency also depends on two covariates: $Z_1=X_1$ and $Z_2$ is independent of the previous ones and normally distributed with mean zero and standard deviation $0.3$. We generate the cure status $B$ as a Bernoulli random variable with success probability  $\phi(\gamma,x)=1/(1+\exp(-\gamma_1-\gamma_2x_1-\gamma_3x_2))$. The survival times for the uncured observations are generated according to a Weibull proportional hazards model
\[
S_u(t|z)=\exp\left(-\mu t^\rho\exp(\beta_1z_1+\beta_2z_2)\right),
\]
and are truncated at $\tau_0$ for $\rho=1.75$ and  $\mu=1.5$. The censoring times are independent from $(T,X,Z)$. They are generated from the exponential distribution with parameter $\lambda_C$ and are truncated at $\tau$.
The mismeasured covariate is $Z_2$, i.e.  we only observe $W=Z_2+U$, where $U\sim N(0,v^2)$ and  $v\in\{0.1, 0.2\}$ corresponding to {small and large} error settings respectively.

\textit{Model 4.} The incidence depends on one covariate  $X$ which is a standard normal random variable.  The latency depends on two covariates: $Z_1=X$ and $Z_2$ is independent of $X$ and uniformly distributed on $[-1,1]$. We generate the cure status $B$ as a Bernoulli random variable with success probability  $\phi(\gamma,x)=1/(1+\exp(-\gamma_1-\gamma_2x))$. The survival times for the uncured observations are generated according to a Weibull proportional hazards model
\begin{equation}
\label{eqn:S_u}
S_u(t|z)=\exp\left(-\mu t^\rho\exp(\beta_1z_1+\beta_2z_2)\right),
\end{equation}
and are truncated at $\tau_0$ for $\rho=1.75$ and  $\mu=1.5$. The censoring times are independent of {the vector $(X,Z,T)$}. They are generated from the exponential distribution with parameter $\lambda_C$ and are truncated at $\tau$.
Instead of $X$ and $Z_2$ we observe $W_1=X+U_1$ and $W_2=Z_2+U_2$, where the error terms $U_1\sim N(0,v_1^2)$ and $U_2\sim N(0,v_2^2)$ are independent.
We consider $(v_1,v_2)=(0.35,0.2)$ and $(v_1,v_2)=(0.7,0.4)$ corresponding to  {small and large} error settings respectively.

\textit{Model 5.} The incidence depends on one covariate  $X$ which is a standard normal random variable.  The latency depends on two correlated covariates: $Z_1=X$ and $Z_2=-X+N$, where $N$ is a normal random variable with mean zero and standard deviation $0.5$ independent of $X$. We generate the cure status $B$ as a Bernoulli random variable with success probability  $\phi(\gamma,x)=1/(1+\exp(-\gamma_1-\gamma_2x))$. The survival times for the uncured observations are generated according to the Weibull proportional hazards model in~\eqref{eqn:S_u}
and are truncated at $\tau_0$ for $\rho=1.75$ and  $\mu=1.5$. The censoring times are independent of {the vector $(X,Z,T)$}. They are generated from the exponential distribution with parameter $\lambda_C$ and are truncated at $\tau$.
The covariate $Z_2$ is measured with error, i.e. instead of $Z_2$ we observe $W=Z_2+U$, where  $U\sim N(0,v^2)$ is independent of the previous variables.
We consider $v=0.39$ and $v=0.78$ corresponding to {small and large} error settings respectively.

\begin{table}[h]
	\centering
	\scalebox{0.85}{
		\begin{tabular}{ccccccccc}
			Model &  Scenario &$\gamma_0$ &$\beta_0$ & $\lambda_C$ & $(\tau_0,\tau)$ & Cure  & Cens.   & Plateau\\
			&  &  & &&& rate &  rate & \\[2pt]
			& & & & &   &&&\\[-8pt]
			& $1$ & $(1.3,1,0.4)$ & $(0.8,0.3)$ &$0.33$ & $(4,6)$ &  $20\%$ & $35\%$& $9\%$\\
			$2$ & $2$ & $(1.1,1.3,-0.3)$ & $(2,-0.8)$ &$0.08$ & $(10,12)$ &  $30\%$ & $35\%$& $19\%$\\
			& $3$ & $(-0.5,1.5,1)$ & $(0.8,0.3)$ &$0.4$ & $(4,6)$ &  $50\%$ & $60\%$& $22\%$\\
			& & & & &   &&&\\[-8pt]
			& $1$ & $(1.3,1,0.4)$ & $(1.5,0.5)$ &$0.3$ & $(6,8)$ &  $20\%$ & $35\%$& $7\%$\\
			$3$ & $2$ & $(1.1,1.3,-0.3)$ & $(1,-1)$ &$0.1$ & $(6,8)$ &  $30\%$ & $35\%$& $22\%$\\
			& $3$ & $(-0.5,1.5,1)$ & $(0.5,1.5)$ &$0.3$ & $(6,8)$ &  $50\%$ & $60\%$& $24\%$\\
			& & & & &   &&&\\[-8pt]
			& $1$ & $(1.4,0.5)$ & $(0.5,0.1)$ &$0.3$ & $(5,7)$ &  $20\%$ & $35\%$& $9\%$\\
			$4$ & $2$ & $(1.4,2)$ & $(0.1,0.5)$ &$0.12$ & $(5,7)$ &  $30\%$ & $35\%$& $22\%$\\
			& $3$ & $(0.-2)$ & $(-1.5,0.5)$ &$0.5$ & $(5,7)$ &  $50\%$ & $60\%$& $14\%$\\
			& & & & &   &&&\\[-8pt]
			& $1$ & $(1.4,0.5)$ & $(0.5,0.1)$ &$0.3$ & $(4,6)$ &  $20\%$ & $35\%$& $10\%$\\
			$5$ & $2$ & $(1.4,2)$ & $(0.1,-0.5)$ &$0.13$ & $(4,6)$ &  $30\%$ & $35\%$& $21\%$\\
			& $3$ & $(0,2)$ & $(1,-1)$ &$0.5$ & $(6,8)$ &  $50\%$ & $60\%$& $12\%$\\
	\end{tabular}}
	\caption{Parameter values and model characteristics for each scenario in Models 2-5.}
	\label{tab:models_2-5}
\end{table}
\begin{table}[h]
	\scalebox{0.85}{
		\begin{tabular}{ccrrrrrrrrrrrr}
		Mod/	&&\multicolumn{3}{c}{naive - 1}&\multicolumn{3}{c}{naive - 2}&\multicolumn{3}{c}{simex - 1}&\multicolumn{3}{c}{simex - 2}\\
			Scen./$v$ 	& Par. &  Bias & Var. & MSE & Bias & Var. & MSE & Bias & Var. & MSE&Bias & Var. & MSE\\[2pt]
					& & & & & & & & & && &&\\[-8pt]
		$2/1 $ & $\gamma_1 $ & $4.8 $  & $12.0 $ & $12.2  $ & $2.7  $ & $11.6 $ & $11.6 $ & $  5.9$ & $12.6 $ & $ 13.0$ & $4.5 $ & $13.0 $ & $13.2 $\\
	$v$=$0.2$	& $\gamma_2 $ & $-6.3 $  & $18.3 $ & $ 18.7 $ & $ -11.3 $ & $ 16.2$ & $17.5 $ & $ 4.7 $ & $ 24.6$ & $ 24.8$ & $-1.0 $ & $ 23.9$ & $23.9 $\\
		& $\gamma_3$ & $ -1.1$  & $ 25.0$ & $25.0  $ & $ -1.7 $ & $ 25.3$ & $ 25.3$ & $-0.7  $ & $ 25.4$ & $25.4 $ & $ -2.2$ & $ 29.4$ & $29.5 $\\
		& $\beta_1 $ & $-8.3 $  & $2.9 $ & $ 3.6 $ & $ -8.1 $ & $ 2.9$ & $ 3.6$ & $ 1.0 $ & $4.0 $ & $ 4.0$ & $1.2 $ & $ 4.0$ & $ 4.0$\\
		& $\beta_2 $ & $-0.2 $  & $ 4.2$ & $  4.2$ & $  -0.2$ & $ 4.1$ & $4.1 $ & $ 0.4 $ & $ 4.3$ & $ 4.3$ & $0.5 $ & $ 4.3$ & $4.3 $\\
		& & & & & & & & & && &&\\[-8pt]
		$2/1 $ & $\gamma_1 $ & $ 2.3$  & $11.8 $ & $ 11.8 $ & $ 0.1 $ & $ 11.7$ & $11.7 $ & $4.4  $ & $ 13.1$ & $13.3 $ & $ 2.6$ & $14.8 $ & $14.9 $\\
		$v$=$0.4$& $\gamma_2 $ & $ -30.3$  & $ 13.4$ & $22.5  $ & $ -34.4 $ & $ 12.1$ & $ 24.0$ & $ -6.0 $ & $ 27.0$ & $27.3 $ & $-11.4 $ & $28.5 $ & $ 29.8$\\
		& $\gamma_3$ & $-1.3 $  & $25.0 $ & $25.0  $ & $-2.2  $ & $ 24.7$ & $24.8 $ & $-0.6  $ & $ 26.3$ & $ 26.3$ & $ -1.7$ & $ 30.1$ & $30.1 $\\
		& $\beta_1 $ & $ -27.2$  & $ 2.1$ & $  9.5$ & $ -27.1 $ & $2.0 $ & $ 9.4$ & $ -7.9 $ & $ 4.3$ & $ 5.0$ & $ -7.7$ & $4.3 $ & $ 4.9$\\
		& $\beta_2 $ & $-1.3 $  & $4.2 $ & $ 4.2 $ & $-1.3  $ & $ 4.2$ & $ 4.2$ & $ 0.0 $ & $ 4.6$ & $ 4.6$ & $ 0.0$ & $ 4.6$ & $ 4.6$\\
			& & & & & & & & & && &&\\[-8pt]
		$3/3$ & $\gamma_1 $  & $-1.8 $  & $7.4 $ & $7.5  $ & $ -1.5 $ & $ 7.5$ & $ 7.6$ & $ -1.8 $ & $7.4 $ & $7.5 $ & $-1.5 $ & $ 7.5$ & $ 7.6$\\
	$v$=$0.1$	& $\gamma_2 $ & $ 5.0$  & $13.4 $ & $ 13.6 $ & $ -2.8$ & $13.6 $ & $ 13.7$ & $4.9  $ & $ 13.4$ & $13.6 $ & $-2.8 $ & $13.6 $ & $ 13.7$\\
		& $\gamma_3$ & $ 3.9$  & $15.2 $ & $ 15.4 $ & $2.0  $ & $15.9 $ & $16.0 $ & $ 3.9 $ & $ 15.2$ & $ 15.4$ & $ 2.0$ & $ 15.9$ & $ 16.0$\\
		& $\beta_1 $ & $-0.3 $  & $ 6.4$ & $  6.4$ & $ 0.4 $ & $ 6.3$ & $6.3 $ & $ 0.3 $ & $ 6.5$ & $6.5 $ & $ 1.0$ & $6.4 $ & $6.4 $\\
		& $\beta_2 $ & $-18.7 $  & $17.1 $ & $ 20.5 $ & $ -19.1 $ & $ 17.0$ & $20.7 $ & $ -3.2 $ & $ 22.7$ & $22.8 $ & $-3.6 $ & $22.7 $ & $22.8 $\\
		& & & & & & & & & && &&\\[-8pt]
		$3/3 $ & $\gamma_1 $ & $-1.8 $  & $7.4 $ & $ 7.5 $ & $ -1.5 $ & $ 7.5$ & $ 7.6$ & $  -1.8$ & $7.5 $ & $ 7.5$ & $ -1.5$ & $7.5 $ & $ 7.6$\\
	$v$=$0.2$	& $\gamma_2 $ & $5.0 $  & $ 13.5$ & $ 13.7 $ & $ -2.8 $ & $13.6 $ & $ 13.7$ & $  5.0$ & $13.5 $ & $13.7 $ & $ -2.8$ & $13.6 $ & $13.7 $\\
		& $\gamma_3$ & $3.9 $  & $ 15.2$ & $ 15.3 $ & $ 2.0 $ & $ 15.9$ & $15.9 $ & $4.0  $ & $15.2 $ & $15.3 $ & $2.0 $ & $15.9 $ & $ 15.9$\\
		& $\beta_1 $ & $-1.6 $  & $6.4 $ & $ 6.4 $ & $ -0.9 $ & $6.3 $ & $6.3 $ & $-0.4  $ & $6.8 $ & $ 6.8$ & $0.3 $ & $ 6.7$ & $6.7 $\\
		& $\beta_2 $ & $-51.6 $  & $13.1 $ & $ 39.7 $ & $ -52.0 $ & $13.0 $ & $40.0 $ & $-17.8  $ & $ 26.9$ & $30.0 $ & $ -18.2$ & $26.9 $ & $30.2 $\\
			& & & & & & & & & && &&\\[-8pt]
		$4/2$& $\gamma_1 $ & $-5.1 $  & $6.4 $ & $ 6.6 $ & $ -9.7 $ & $5.9 $ & $ 6.9$ & $4.3  $ & $ 8.7$ & $ 8.9$ & $ -1.7$ & $8.2 $ & $ 8.2$\\
		$v_1$=$0.35$& $\gamma_2 $ & $-27.5 $  & $ 8.8$ & $16.4  $ & $ -39.6 $ & $8.0 $ & $23.7 $ & $ 4.5 $ & $ 15.9$ & $ 16.1$ & $ -11.6$ & $15.1 $ & $ 16.5$\\
		$v_2$=$0.2$&  $\beta_1 $ & $-1.4 $  & $1.2 $ & $1.2  $ & $ -1.3 $ & $1.2 $ & $ 1.2$ & $ 0.1 $ & $1.7 $ & $1.7 $ & $0.3 $ & $ 1.7$ & $1.7 $\\
		& $\beta_2 $ & $-3.6 $  & $ 2.6$ & $2.8  $ & $  -3.6$ & $2.6 $ & $2.8 $ & $ 2.0 $ & $3.5 $ & $ 3.6$ & $2.0 $ & $3.5 $ & $ 3.6$\\
		& & & & & & & & & && &&\\[-8pt]
		$4/2$ & $\gamma_1 $ & $ -21.6$  & $ 4.9$ & $ 9.6 $ & $ -24.7 $ & $ 4.6$ & $ 10.7$ & $ -7.2 $ & $8.1 $ & $ 8.6$ & $ -11.9$ & $ 7.5$ & $8.9 $\\
		$v_1$=$0.7$& $\gamma_2 $ & $-85.1 $  & $4.5 $ & $  76.9$ & $-93.0  $ & $4.2 $ & $90.7 $ & $ -33.7 $ & $ 12.6$ & $ 23.9$ & $-46.0 $ & $11.9 $ & $33.1 $\\
		$v_2$=$0.4$&  $\beta_1 $ & $-4.4 $  & $0.8 $ & $ 1.0 $ & $ -4.2 $ & $0.8 $ & $ 1.0$ & $ -1.8 $ & $ 1.8$ & $ 1.8$ & $-1.6 $ & $1.7 $ & $1.8 $\\
		& $\beta_2 $ & $15.5 $  & $ 1.9$ & $4.3  $ & $ -15.5 $ & $ 1.9$ & $4.3 $ & $ -3.2 $ & $3.9 $ & $4.0 $ & $ -3.2$ & $3.9 $ & $4.0 $\\
		& & & & & & & & & && &&\\[-8pt]
		$5/3$ & $\gamma_1 $ & $1.2 $  & $ 7.0$ & $ 7.0 $ & $ -0.8 $ & $ 6.6$ & $ 6.6$ & $ 1.0 $ & $ 6.9$ & $6.9 $ & $ -0.8$ & $6.6 $ & $6.6 $\\
	$v$=$0.39$	& $\gamma_2 $ & $11.4 $  & $20.3 $ & $ 21.6 $ & $3.8  $ & $ 20.9$ & $ 21.1$ & $ 11.7 $ & $20.5 $ & $21.8 $ & $3.8 $ & $ 20.9$ & $ 21.1$\\
		& $\beta_1 $ & $ 34.0$  & $8.8 $ & $ 20.3 $ & $  34.5$ & $ 8.8$ & $20.7 $ & $14.1  $ & $ 11.9$ & $ 13.9$ & $14.8 $ & $11.9 $ & $14.1 $\\
		& $\beta_2 $ & $38.4 $  & $4.7 $ & $  19.4$ & $ 38.8 $ & $ 4.7$ & $19.7 $ & $ 13.1 $ & $10.7 $ & $ 12.4$ & $13.6 $ & $ 10.7$ & $12.5 $\\
		& & & & & & & & & && &&\\[-8pt]
		$5/3 $ & $\gamma_1 $ & $1.5 $  & $ 7.2$ & $  7.2$ & $ -0.8 $ & $ 6.6$ & $6.6 $ & $1.3  $ & $7.2 $ & $7.2 $ & $-0.8 $ & $6.6 $ & $6.6 $\\
	$v$=$0.78$	& $\gamma_2 $ & $11.0 $  & $ 20.5$ & $21.7  $ & $  3.8$ & $ 20.9$ & $ 21.1$ & $ 11.2 $ & $20.7 $ & $ 22.0$ & $ 3.8$ & $ 20.9$ & $ 21.1$\\
		& $\beta_1 $ & $60.3 $  & $7.7 $ & $ 44.0 $ & $ 60.6 $ & $7.7 $ & $44.4 $ & $ 45.5 $ & $ 10.1$ & $30.8 $ & $ 46.0$ & $ 10.1$ & $31.2 $\\
		& $\beta_2 $ & $ 72.0$  & $ 2.1$ & $ 54.0 $ & $ 72.2 $ & $ 2.1$ & $54.2 $ & $ 53.4 $ & $6.6 $ & $35.2 $ & $ 53.7$ & $6.6 $ & $35.5 $\\
			\end{tabular}}
	\caption{Bias, variance and MSE of $\hat\gamma$ and $\hat\beta$ for the naive and simex method  based on the maximum likelihood  (1) or the  presmoothing  (2) approach in Models 2-5 ($n=200$). The first column gives the model, scenario and the standard deviation of the measurement error. All numbers were multiplied by $100$.}
	\label{tab:results_2}
\end{table}

For the four models, various choices  of the parameters $\gamma$, $\beta$, $\lambda_C$ and $(\tau_0,\tau)$ are considered, in such a way that we obtain three scenarios for the cure rate ($20\%$, $30\%$ and $50\%$) and different levels of censoring (see Table \ref{tab:models_2-5}).  The sample size is fixed at $n=200$, while the variance of the measurement error is chosen as described in each model, corresponding to a ratio between the standard deviation of the error and the standard deviation of the covariate equal to $0.35$ and $0.7$. Some of the results  are given in Table~\ref{tab:results_2} and the rest can be found in Tables~\ref{tab:results_3}-\ref{tab:results_4} in the Supplementary Material.

Once more we observe that the maximum likelihood estimator and the estimator based on presmoothing give comparable results for both the naive and the simex method. As expected, the measurement error mainly affects the estimators of the  coefficients corresponding to the mismeasured covariates. However,
the measurement error induces bias also on variables correlated to the mismeasured covariate within the same component. For example in Model 5, the measurement error of $Z_2$ leads to biased estimators for $\beta_1$ and $\beta_2$, but does not affect the estimation of $\gamma_2$ even though $Z_1=X$. In all settings, the simex method corrects for the bias due to the measurement error. Nevertheless, in terms of mean squared error, the naive approach is still preferred  when the measurement error is small and the absolute value of the coefficient corresponding to the standardized covariate is small (the covariate has a weak effect on cure or survival). On the contrary, a strong effect (large coefficient) and a large measurement error favour the use of the simex method. 
\subsection{Robustness of the method}
Here we investigate the robustness of the simex approach with respect to the choice of the extrapolation function, misspecification of the  error distribution and of the error standard deviation. We focus on Model 2, where the mismeasured covariate is $X_1=Z_1$ affecting both the cure probability and the survival. The sample size is $n=200$ and the error standard deviation is $v=0.2$ or $v=0.4$. 

In addition to the quadratic extrapolant used in Table~\ref{tab:results_2}, we consider also a linear and a cubic extrapolant. Results in Table~\ref{tab:results_extrapolation} show that, as the order of the extrapolation function increases, the difference between the maximum likelihood estimators and the estimators based on presmoothing becomes more significant. In particular, it favours the first method over the latter one mainly due to a smaller variance.  As expected, the choice of the extrapolation function has stronger effect on the coefficients corresponding to the mismeasured covariates and when the error is large.  For $v=0.4$, the bias decreases as the extrapolation order increases while there is no clear conclusion when $v$ is small. In terms of mean squared error, linear extrapolation is preferred when the measurement error variance is low or more in general in situations where the naive method would do better than the simex approach. In cases where simex outperforms the naive estimators, the quadratic extrapolant seems to be the best choice. 
\begin{table}[h]
	\centering
	\scalebox{0.85}{
		\begin{tabular}{ccrrrrrrrrrrrr}
			&\multicolumn{7}{c}{}&\multicolumn{6}{c}{}\\[-10pt]
			&&\multicolumn{6}{c}{$v=0.2$}&\multicolumn{6}{c}{$v=0.4$}\\
			&&\multicolumn{3}{c}{simex - 1}&\multicolumn{3}{c}{simex - 2}&\multicolumn{3}{c}{simex - 1}&\multicolumn{3}{c}{simex - 2}\\
			& Par. &  Bias & Var. & MSE & Bias & Var. & MSE & Bias & Var. & MSE&Bias & Var. & MSE\\[2pt]
			& & & & & & & & & && &&\\[-5pt]
			linear & $\gamma_1 $ & $ 5.6$  & $12.3 $ & $ 12.6  $ & $ 3.6 $ & $ 12.0$ & $ 12.1 $ & $  3.3 $ & $  12.2$ & $  12.3$ & $1.3  $ & $ 12.3 $ & $12.3  $\\
			& $\gamma_2 $ & $1.9 $  & $ 21.9$ & $ 22.0  $ & $ -3.5 $ & $ 19.8$ & $20.0  $ & $ -18.2  $ & $18.7  $ & $22.0  $ & $ -22.9 $ & $ 17.0 $ & $ 22.3 $\\
			& $\gamma_3 $ & $ -0.9$  & $25.4 $ & $ 25.4  $ & $ -1.6 $ & $ 26.0$ & $26.1  $ & $ -1.0  $ & $ 25.7 $ & $ 25.7 $ & $ -2.2 $ & $25.5  $ & $ 25.5 $\\
			& $\beta_1 $ & $-1.7 $  & $3.6 $ & $ 3.7  $ & $-1.5  $ & $3.6 $ & $ 3.6 $ & $ -17.9 $ & $  2.9$ & $ 6.2 $& $  -17.8 $ & $ 2.9 $ & $ 6.0 $ \\
			& $\beta_2 $ & $0.3 $  & $4.3 $ & $  4.3 $ & $  0.3$ & $ 4.2$ & $ 4.2 $ & $ -0.7  $ & $ 4.3 $ & $  4.4$ & $  -0.6$ & $ 4.3 $ & $ 4.3 $\\
			& & & & & & & & & && &&\\[-5pt]
			cubic & $\gamma_1 $ & $6.5 $  & $ 12.7$ & $ 13.1  $ & $5.3  $ & $ 16.7$ & $ 16.9 $ & $ 5.6  $ & $ 13.5 $ & $ 13.8 $ & $4.1  $ & $ 22.6 $ & $ 22.7 $\\
			& $\gamma_2 $ & $5.9 $  & $28.1 $ & $ 28.4  $ & $ -0.5 $ & $ 34.2$ & $34.2  $ & $0.5   $ & $ 35.4 $ & $35.4  $ & $ -4.4 $ & $ 54.4 $ & $ 54.6 $\\
			& $\gamma_3 $ & $-1.1 $  & $ 25.3$ & $ 25.3  $ & $ -2.2 $ & $ 39.2$ & $  39.2$ & $  -0.8 $ & $ 26.6 $ & $ 26.6 $ & $ -2.8 $ & $ 46.0 $ & $ 46.1 $\\
			& $\beta_1 $ & $ 2.3$  & $5.0 $ & $5.0   $ & $ 2.6 $ & $4.9 $ & $ 5.0 $ & $  -2.4 $ & $ 6.3 $ & $ 6.4 $ & $-2.2  $ & $ 6.3 $ & $ 6.4 $\\
			& $\beta_2 $ & $0.2 $  & $ 4.5$ & $ 4.5  $ & $  0.3$ & $4.4 $ & $ 4.4 $ & $-0.1   $ & $ 5.0 $ & $ 5.0 $ & $ 0.0 $ & $ 5.0 $ & $5.0  $\\
	\end{tabular}}
	\caption{Bias, variance and MSE of $\hat\gamma$ and $\hat\beta$ for the simex method  based on the maximum likelihood  (1) or the  presmoothing  (2) approach with three different extrapolation functions. All numbers were multiplied by $100$.}
	\label{tab:results_extrapolation}
\end{table}

To understand what happens if the error distribution is misspecified we generate the measurement error from three other distributions: a uniform distribution  $U\sim \mathrm{Unif}(-a,a)$, a Student-t distribution with $k$ degrees of freedom $a^{-1}U\sim t_k$ and a chi-squared distribution with $k$ degrees of freedom $a^{-1}U+k\sim \chi^2_k$. The constant $a$ is chosen in such a way that the standard deviation of $U$ is $v=0.2$ or $v=0.4$. In all three cases we still use the Gaussian distribution in the simex procedure. Results are given in Table~\ref{tab:results_distribution}. We observe that, when the true distribution is uniform or Student-t, the method still behaves quite well and there is  little impact on the estimators.  However, when the true distribution is not symmetric ($\chi^2$) there is a significant increase in mean squared error, in particular for large $v$.  
\begin{table}[h]
	\centering
	\scalebox{0.85}{
		\begin{tabular}{ccrrrrrrrrrrrr}
			&\multicolumn{7}{c}{}&\multicolumn{6}{c}{}\\[-10pt]
			&&\multicolumn{6}{c}{$v=0.2$}&\multicolumn{6}{c}{$v=0.4$}\\
			&&\multicolumn{3}{c}{simex - 1}&\multicolumn{3}{c}{simex - 2}&\multicolumn{3}{c}{simex - 1}&\multicolumn{3}{c}{simex - 2}\\
			& Par. &  Bias & Var. & MSE & Bias & Var. & MSE & Bias & Var. & MSE&Bias & Var. & MSE\\[2pt]
			& & & & & & & & & && &&\\[-5pt]
			t-distr. & $\gamma_1 $ & $6.3 $  & $ 12.4$ & $  12.7 $ & $  3.9$ & $13.6 $ & $ 13.7 $ & $ 5.3  $ & $ 12.7 $ & $ 12.9 $ & $ 2.5 $ & $ 14.3 $ & $ 14.4 $\\
			& $\gamma_2 $ & $ 5.9$  & $ 23.2$ & $23.6   $ & $0.5  $ & $25.8 $ & $25.9  $ & $  -1.3 $ & $ 27.0 $ & $27.1  $ & $ -5.2 $ & $ 26.5 $ & $  26.7$\\
			& $\gamma_3 $ & $-1.5 $  & $25.3 $ & $25.4   $ & $ -0.1 $ & $ 31.2$ & $ 31.2 $ & $ -2.1  $ & $  26.0$ & $ 26.0 $ & $ -1.6 $ & $ 33.8 $ & $ 33.8 $\\
			& $\beta_1 $ & $ 0.1$  & $ 4.1$ & $  4.1 $ & $ 0.3 $ & $4.0 $ & $4.0  $ & $ -8.7  $ & $4.8  $ & $5.6  $ & $ -8.5 $ & $ 4.8 $ & $ 5.5 $\\
			& $\beta_2 $ & $0.5 $  & $ 4.3$ & $  4.3 $ & $0.3  $ & $4.3 $ & $4.3  $ & $  0.2 $ & $  4.6$ & $ 4.6 $ & $ 0.1 $ & $ 4.6 $ & $ 4.6 $\\
			& & & & & & & & & && &&\\[-5pt]
			Unif. & $\gamma_1 $ & $6.2 $  & $ 12.4$ & $ 12.7  $ & $4.9  $ & $ 14.0$ & $ 14.2 $ & $ 5.1  $ & $ 12.6 $ & $ 12.9 $ & $ 4.5 $ & $ 15.0 $ & $ 15.2 $\\
			& $\gamma_2 $ & $ 4.9$  & $22.4 $ & $  22.6 $ & $ -1.5 $ & $ 24.5$ & $ 24.5 $ & $ -5.8  $ & $ 23.3 $ & $  23.6$ & $ -11.2 $ & $ 27.1 $ & $28.3  $\\
			& $\gamma_3 $ & $-1.2 $  & $25.1 $ & $ 25.1  $ & $ -0.2 $ & $ 29.4$ & $  29.4$ & $ -1.5  $ & $ 25.4 $ & $ 25.4 $ & $ -0.8 $ & $ 29.0 $ & $ 29.0 $\\
			& $\beta_1 $ & $0.1 $  & $4.0 $ & $  4.0 $ & $  0.4$ & $4.0 $ & $ 4.0 $ & $  -9.5 $ & $ 4.3 $ & $ 5.2 $ & $ -9.2 $ & $ 4.2 $ & $ 5.0 $\\
			& $\beta_2 $ & $ 0.3$  & $4.3 $ & $  4.3 $ & $ 0.2 $ & $ 4.4$ & $ 4.4 $ & $ -0.2  $ & $ 4.7 $& $  4.7$ & $ -0.3 $ & $ 4.6 $ & $ 4.6 $ \\
			& & & & & & & & & && &&\\[-5pt]
			$\chi^2 $& $\gamma_1 $ & $6.9 $  & $ 12.7$ & $  13.2 $ & $  3.2$ & $ 13.1$ & $13.2  $ & $8.2   $ & $ 14.6 $ & $ 15.3 $ & $  4.1$ & $ 14.8 $ & $ 14.9 $\\
			& $\gamma_2 $ & $ 6.8$  & $25.6$ & $ 26.1  $ & $-2.4  $ & $25.6 $ & $ 25.7 $ & $ 5.6  $ & $ 38.6 $ & $38.9  $ & $ -13.5 $ & $ 33.8 $ & $ 35.7 $\\
			& $\gamma_3 $ & $-1.5 $  & $ 25.1$ & $ 25.1  $ & $  -0.5$ & $27.7 $ & $27.7  $ & $-2.0   $ & $25.9  $ & $ 25.9 $ & $  -1.4$ & $  33.1$ & $ 33.1 $\\
			& $\beta_1 $ & $ -0.5$  & $4.5 $ & $ 4.5  $ & $-0.1  $ & $ 4.4$  & $ 4.4  $& $  -12.9$ & $ 6.0 $ & $ 7.7$ &$-11.8 $ & $5.9  $ & $ 7.3 $ \\
			& $\beta_2 $ & $0.2 $  & $4.3 $ & $ 4.3  $ & $  0.1$ & $ 4.4$ & $ 4.4 $ & $ -0.5  $ & $ 4.8 $ & $ 4.8 $ & $ -0.5 $ & $ 4.8 $ & $4.8  $\\
	\end{tabular}}
	\caption{Bias, variance and MSE of $\hat\gamma$ and $\hat\beta$ for the simex method based on the maximum likelihood  (1) or the  presmoothing  (2) approach when the error distribution is misspecified. All numbers were multiplied by $100$.}
	\label{tab:results_distribution}
\end{table}

Finally we investigate the effect of error variance misspecification. We simulate the error from a normal distribution with standard deviation $v=0.2$ and $v=0.4$ but in the estimation process the variance is misspecified $v_E\in\{v-0.1,v+0.1\}$. Results reported in Table~\ref{tab:results_variance} show that the misspecification affects estimation of all the parameters but the difference is larger for those that correspond to the mismeasured covariates. As expected, increasing the specified variance $v_E^2$ leads to an increased variance of the simex  estimators. For small $v$, the lowest bias is obtained when $v$ is correctly specified while for large $v$, the bias decreases as the specified variance increases. In terms of mean squared error, in situations where simex performs worse than the naive approach underspecifying the variance works better. On the other hand, when simex outperforms the naive estimators, overspecifying the error variance is preferred over underspecification. 

\begin{table}[h]
	\centering
	\scalebox{0.85}{
		\begin{tabular}{ccrrrrrrrrrrrr}
			&\multicolumn{7}{c}{}&\multicolumn{6}{c}{}\\[-10pt]
			&&\multicolumn{6}{c}{$v=0.2$}&\multicolumn{6}{c}{$v=0.4$}\\
			&&\multicolumn{3}{c}{simex - 1}&\multicolumn{3}{c}{simex - 2}&\multicolumn{3}{c}{simex - 1}&\multicolumn{3}{c}{simex - 2}\\
			$v_E$& Par. &  Bias & Var. & MSE & Bias & Var. & MSE & Bias & Var. & MSE&Bias & Var. & MSE\\[2pt]
			& & & & & & & & & && &&\\[-10pt]
			$v-0.1$& $\gamma_1 $ & $5.0 $  & $ 12.2$ & $ 12.4  $ & $ 3.3 $ & $ 12.1$ & $ 12.2 $ & $  3.5 $ & $ 12.6 $ & $ 12.7 $ & $1.6 $ & $  13.7$ & $ 13.8 $\\
			& $\gamma_2 $ & $ -3.7$  & $ 19.9$ & $20.0   $ & $  -8.9$ & $ 18.4$ & $ 19.2 $ & $  -16.4 $ & $20.9  $ & $ 23.6 $ & $-21.2 $ & $ 22.6 $ & $ 27.1 $\\
			& $\gamma_3 $ & $ -1.0$  & $ 25.1$ & $ 25.1  $ & $ -2.2 $ & $ 27.3$ & $  27.4$ & $ -0.9 $ & $ 25.7 $ & $ 25.7 $ & $ -2.1 $ & $ 28.6 $ & $ 28.6$\\
			& $\beta_1 $ & $-6.1 $  & $ 3.2$ & $3.6   $ & $ -5.9 $ & $3.1 $ & $ 3.5 $ & $ -16.1  $ & $  3.3$ & $  5.9$ & $ -15.9 $ & $3.3 $ & $ 5.8$\\
			& $\beta_2 $ & $0 .0$  & $ 4.2$ & $  4.2 $ & $0.0  $ & $ 4.2$ & $ 4.2 $ & $ -0.6 $ & $ 4.4 $ & $ 4.4 $ & $ -0.5 $ & $ 4.4 $ & $4.4  $\\
			& & & & & & & & & && &&\\[-5pt]
			$v+0.1$ & $\gamma_1 $ & $7.4 $  & $ 13.4$ & $  13.9 $ & $ 6.0 $ & $ 14.0$ & $ 14.4 $ & $  5.5 $ & $  13.8$ & $ 14.1 $ & $ 3.9 $ & $ 15.6 $ & $ 15.8 $\\
			& $\gamma_2 $ & $18.6 $  & $ 32.7$ & $  36.1 $ & $ 12.2 $ & $32.1 $ & $ 33.6 $ & $ 5.5  $ & $34.2  $ & $ 34.5 $ & $ 0.1 $ & $ 35.7 $ & $ 35.7$\\
			& $\gamma_3 $ & $ -0.4$  & $26.1 $ & $ 26.1  $ & $  -1.5$ & $ 31.3$ & $  31.3$ & $ -0.3  $ & $ 27.1 $ & $  27.1$ & $  -1.7$ & $ 31.4 $ & $31.4  $\\
			& $\beta_1 $ & $ 12.4$  & $5.5 $ & $7.1   $ & $ 12.6 $ & $ 5.5$ & $7.1  $ & $ 1.1  $ & $  5.6$ & $ 5.6$ & $  1.3$ & $  5.5$ & $ 5.5$\\
			& $\beta_2 $ & $1.2 $  & $4.6 $ & $ 4.6  $ & $ 1.3 $ & $ 4.6$ & $ 4.6 $ & $   0.6$ & $  4.9$ & $ 4.9$ & $  0.7$ & $ 4.9$ & $ 4.9 $\\
	\end{tabular}}
	\caption{Bias, variance and MSE of $\hat\gamma$ and $\hat\beta$ for the simex method based on the maximum likelihood  (1) or the  presmoothing  (2) approach when the  error variance is misspecified. All numbers were multiplied by $100$.}
	\label{tab:results_variance}
\end{table}
\section{Application: prostate cancer study}
\label{sec:application}
In this section we illustrate the practical use of the proposed simex procedure for a medical dataset concerning patients with prostate cancer. 
According to the American Cancer Society, prostate cancer is the second most common cancer among American men (after skin cancer) and it is estimated that about 1 man in 9 is diagnosed with prostate cancer during his lifetime. Even though most men diagnosed with prostate cancer do not die from it, it can sometimes be a serious disease.  The 5-year survival rate based on the stage of the cancer at diagnoses  is almost  $100\%$ for localized or regional stage and drops to $31\%$ for distant stage. Among other factors, the prostate-specific antigen (PSA) blood level is a good indicator of the presence of the cancer and is used as a tool to both diagnose and monitor the development of the disease. In most cases, elevated PSA levels indicate a poor prostate cancer prognosis. Even though most studies do not take it in consideration, the PSA measurements are not error-free because of the inaccuracy of the measuring technique and own fluctuations of the PSA levels. Here we try to analyse the effect of PSA on cure probability and survival while accounting for measurement error. 

\begin{figure}
	\centering
	\includegraphics[width=0.48\linewidth]{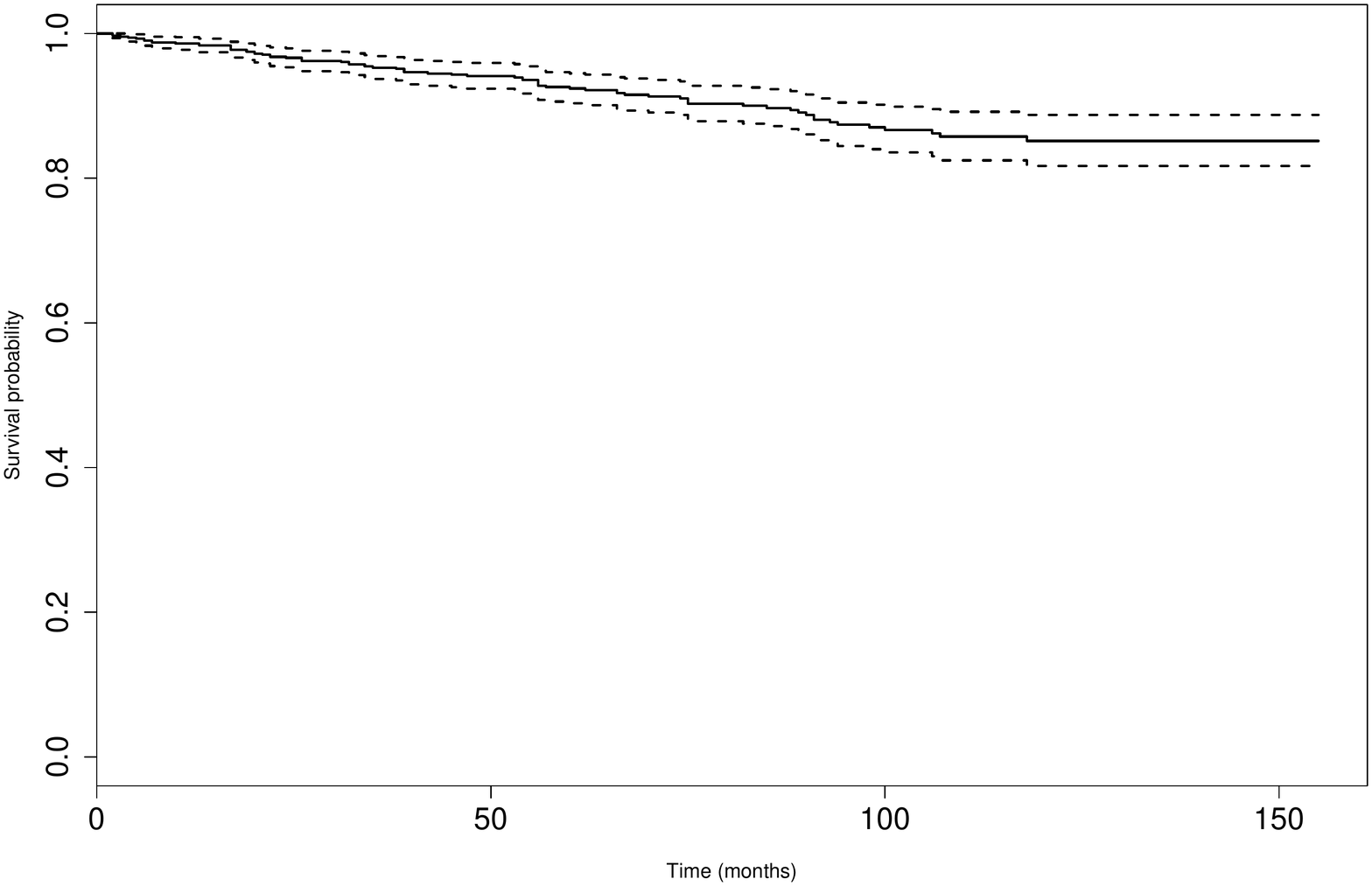}
	\includegraphics[width=0.48\linewidth]{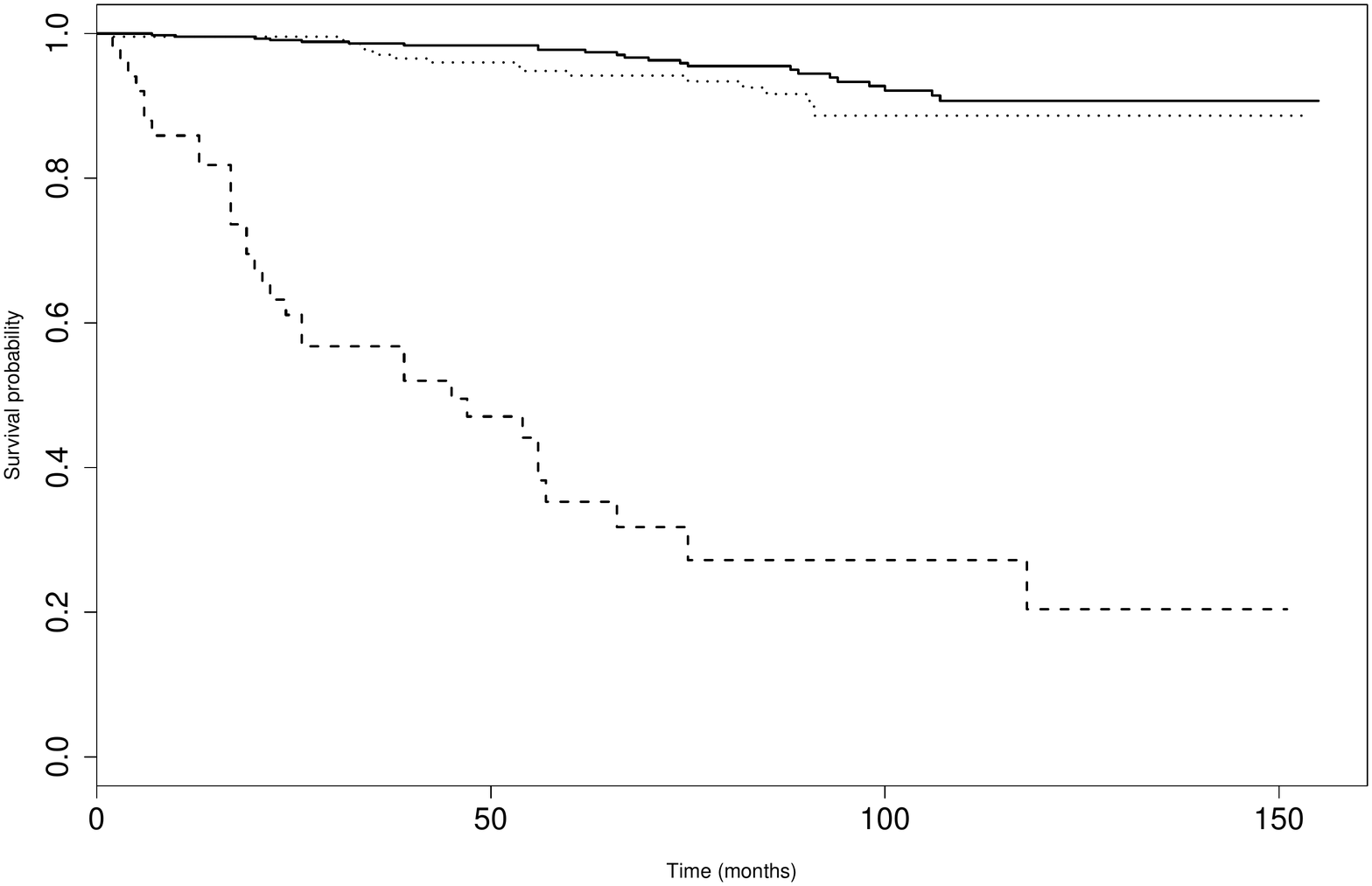}
	\caption{Left panel: Kaplan Meier survival curve for the prostate cancer data. Right panel: Kaplan Meier survival curves  based on cancer stage at diagnosis, localized (solid), regional (dotted) and distant (dashed).}
	\label{fig:KM_prostate}
\end{figure}

We obtain the data from the Surveillance, Epidemiology and End Results (SEER) database, which is a collection of cancer incidence data from population-based cancer registries in the US. We select the database 'Incidence - SEER 18 Regs Research Data' and extract the prostate cancer data for the county of San Bernardino in California during the period $2004-2014$. We restrict to only white patients, aged $35-65$ years old, with stage at diagnosis: localized, regional or distant and  follow-up time greater than zero.  Since a PSA level smaller than $4$ ng/ml of blood is considered as normal and a PSA value between $4$ and $10$ ng/ml is considered as a borderline range, we focus only on patients with PSA level greater than $10$ ng/ml. The event time is death because of prostate cancer.  This cohort consists of $726$ observations out of which $654$ do not experience cancer related death (i.e. around $90\%$ are censored). The follow-up time ranges from $2$ to $155$ months. For most of the patients the cancer has been diagnosed at early stage (localized), while for $228$ of them the stage at diagnosis is `regional' and only for $51$ it is `distant'. The PSA level varies from $10.1$ to $94$ ng/ml, with median  value $15.4$ ng/ml, mean value $21.9$ ng/ml and standard deviation $16$ ng/ml.  We use a logistic/Cox mixture cure model to analyse this dataset and the covariates of interest are the PSA level (continuous variable centered to the mean and measured with error) and stage at diagnosis. The latter one  is classified using two dummy  Bernoulli variables $S_1$ and $S_2$, indicating  distant and regional stage respectively. 
The use of the cure models is justified from the presence of a long plateau containing around $18\%$ of the observations visible in the Kaplan-Meier curve {(\cite{KM})} in Figure \ref{fig:KM_prostate}. Moreover, the Kaplan-Meier curves depending on stage at diagnosis  in Figure~\ref{fig:KM_prostate} confirm that being in the distant stage significantly affects the probability of being cured.
We first estimate the model ignoring the measurement error (`naive') and then we apply the simex procedure {with quadratic extrapolation function} for two levels of measurement error, namely with standard deviation $v=4.8$ and $v=8$, {corresponding to a ratio between the standard deviation of the error and the standard deviation of the covariate equal to $0.3$ and $0.5$ (we considered slightly smaller error than in the simulation setting in order to be closer to real life scenarios). 
} In all three cases we use both the maximum likelihood estimation method and the presmoothing based method. The standard deviations of the estimates are computed through $1000$ bootstrap samples. We consider such a large number of bootstrap samples because we noted that the estimated standard deviation for $\gamma_3$ (distant stage) is not very stable due to  the small sample size of that category. The results are reported in Table~\ref{tab:prostate_cancer}. 

First of all we observe that, independently of the estimation method that we use, the PSA level and being in the distant stage are significant for the cure probability, while only the latter one is significant for survival of uncured patients (at level $5\%$). The positive sign of the coefficients  confirms that high PSA level and distant stage are related to low cure probability and  poor survival.  Note that the estimated coefficient for the PSA value seems very small but it corresponds to a coefficient around $0.5$ for the standardized variable. Given that the sample size is also large, we expect that, if the measurement error is relatively large, the use of the simex procedure would give more accurate results. Moreover, since there is some correlation between the PSA level and the stage of cancer, the measurement error might induce bias also in the other coefficients. For the maximum likelihood estimator, the estimated effect of the PSA level on the cure probability is slightly stronger when  taking into account the measurement error, while the effect of the distant stage is slightly weakened. The opposite happens with the estimation based on presmoothing. To understand what these differences in the estimates mean in practical terms we compute the cure probability for patients with  distant or localized stage and three different PSA levels: $10$ ng/ml, $22$ ng/ml (mean value) and $34$ ng/ml (see Table~\ref{tab:cure_prostate}). {Contrary to our expectation, we see that, in this example, there is not much difference between the naive and the simex approach. We observed in the simulation study that, when the bias induced by the measurement error is large, it is significantly reduced by the simex procedure and otherwise simex has little effect (see for example estimation of $\gamma_2$ in Model 1 and Model 2, Scenario 1 with $v=0.2$  or estimation of $\beta_2$ in Model 5, Scenario 2 with $v_2=0.2$). Hence, we can conclude that in this example, the bias induced by the mismeasured PSA  value is small. This is probably due to the fact that the effect of the PSA value on survival is weak (the absolute  value of its coefficient is small compared to the intercept and the coefficient of $S_1$). The very high cure and censoring rate might also play a  role. On the other hand, correlation between PSA and the stage of cancer would lead to induced bias even for the coefficients corresponding to $S_1$ and $S_2$. From the simulation study (see Model 5) we expect this bias to be of the same order as for the mismeasured covariate. Thus, since here the bias for the coefficient of the PSA value is small, even for the coefficients of $S_1$ and $S_2$ we do not  observe much difference between the naive and simex method. Finally, we find that the estimated cure probabilities are larger when using the estimators based on presmoothing. Based again on the simulation study (cases with small bias), it is more likely that presmoothing behaves better than the maximum likelihood approach.}

\begin{table}
	\centering
	\scalebox{0.85}{
	\begin{tabular}{ccccrrrrrrrr}
		&	& & &\multicolumn{4}{c}{incidence} & & \multicolumn{3}{c}{latency}\\
		&	& &	&& & & & & & & \\[-12pt]
		&	& &&
			 Intercept
	& PSA  & $S_1$ & $S_2$ & &  PSA  & $S_1$ & $S_2$\\[2pt]
		&	&& & && & && & & \\[-10pt]
		&	\multirow{3}{*}{\STAB{\rotatebox[origin=c]{90}{naive-1}}} & estimates &&$-2.2307 $ & $0.0302 $ & $3.2982 $ & $ 0.1021$ & &$ 0.0081 $ & $1.2775 $ & $ 0.6120$\\
		&	&	est. SD && $0.2043 $ & $ 0.0097$ & $ 1.1943$ & $ 0.3802$ & &$ 0.0078 $ & $0.5060 $ & $ 0.3485$\\
		&	&		p-value && $ 0.0000$ & $0.0019 $ & $ 0.0058$ & $0.7883 $ & &$ 0.2999 $ & $0.0116 $ & $0.0790 $\\
		&		&		&	& & && & & & & \\[-10pt]
		&	&		&	& & & & && & & \\[-10pt]
		&\multirow{3}{*}{\STAB{\rotatebox[origin=c]{90}{naive-2}}}	&		estimates && $-2.3050 $ & $0.0293 $ & $3.2373 $ & $ 0.1909$ && $  0.0081$ & $1.2600 $ & $0.5667 $\\
		&	&		est. SD && $0.2221 $ & $0.0084 $ & $ 0.4950$ & $ 0.3780$ && $0.0072  $ & $0.4937 $ & $0.3219 $\\
		&	&		p-value && $ 0.0000$ & $0.0005 $ & $ 0.0000$ & $0.6135 $ && $0.2619  $ & $0.0107 $ & $0.0852 $\\
		&				&&			& & & & & & & \\[-10pt]
		&	&			& & &&& & & & & \\[-10pt]
		\multirow{6}{*}{\STAB{\rotatebox[origin=c]{90}{$v=4.8$}}} & \multirow{3}{*}{\STAB{\rotatebox[origin=c]{90}{simex-1}}}	&		estimates && $ -2.2311$ & $0.0306 $ & $ 3.2927$ & $0.0990 $ && $0.0081  $ & $ 1.2757$ & $0.6151 $\\
		&	&		est. SD && $0.2048 $ & $0.0099 $ & $1.1939 $ & $ 0.3809$ && $ 0.0079 $ & $0.5065 $ & $0.3492 $\\
		&	&		p-value && $0.0000 $ & $0.0019 $ & $ 0.0058$ & $0.7949 $ && $0.3066  $ & $0.0118 $ & $0.0782 $\\
		&			&&		&	& & & & & & & \\[-10pt]
		&	&		&	& && & & & & & \\[-10pt]
		&\multirow{3}{*}{\STAB{\rotatebox[origin=c]{90}{simex-2}}}	&	estimates& & $-2.2757 $ & $0.0281 $ & $ 3.2500$ & $ 0.1779$ && $ 0.0085 $ & $1.2658 $ & $0.5766 $\\
		&	&		est. SD && $0.2937 $ & $0.0106 $ & $ 0.6480$ & $0.5603 $ && $ 0.0073 $ & $ 0.4971$ & $ 0.3405$\\
		&	&		p-value& & $0.0000$ & $0.0083 $ & $0.0000 $ & $0.6992 $ && $ 0.2442 $ & $0.0109 $ & $ 0.0903$\\
		&		&	&		&	& & & & & & & \\[-10pt]
		&	&		&	& & && & & & & \\[-10pt]
		\multirow{6}{*}{\STAB{\rotatebox[origin=c]{90}{$v=8$}}} & \multirow{3}{*}{\STAB{\rotatebox[origin=c]{90}{simex-1}}}	&		estimates && $ -2.2337$ & $ 0.0317$ & $ 3.2857$ & $ 0.0995$ && $0.0081  $ & $1.2836 $ & $ 0.6150$\\
		&	&		est. SD && $0.2046 $ & $0.0099 $ & $1.1957 $ & $0.3814 $ && $ 0.0080 $ & $0.5064 $ & $ 0.3494$\\
		&	&		p-value& & $ 0.0000$ & $0.0014 $ & $0.0060 $ & $ 0.7942$ && $ 0.3092 $ & $00112 $ & $ 0.0784$\\
		&			&	&	&	& & & & & & & \\[-10pt]
		&	&		&	& & & && & & & \\[-10pt]
		&\multirow{3}{*}{\STAB{\rotatebox[origin=c]{90}{simex-2}}}	&	estimates && $-2.2752 $ & $ 0.0285$ & $3.2412 $ & $0.2110 $& & $0.0086  $ & $ 1.2746$ & $ 0.5655$\\
		&	&		est. SD& & $0.2992 $ & $0.0110 $ & $ 0.6687$ & $0.4712 $ && $ 0.0074 $ & $0.4996 $ & $0.3434 $\\
		&	&		p-value& & $ 0.0000$ & $0.0096 $ & $ 0.0000$& $0.6543 $ & &$0.2444 $ & $ 0.0107 $ & $ 0.0997$ \\
		
	\end{tabular}}
	\caption{Coefficient estimates, estimated standard deviations and p-values for the prostate cancer data using the naive and the simex method based on the maximum likelihood  (1) and the presmoothing (2) approach.}
	\label{tab:prostate_cancer}
\end{table}

\begin{table}
	\centering
	\scalebox{0.85}{
	\begin{tabular}{lrrrrrr}
		&	 \multicolumn{3}{c}{`Localized'} & \multicolumn{3}{c}{`Distant'}\\
		&	& 	& &&& \\[-12pt]
		PSA (ng/ml)	& $10$	& $22$& $34$ 	& $10$	& $22$& $34$\\[2pt]
		&	&& & & &  \\[-10pt]
		naive - 1 & $93.0\%$ & $90.3\%$ & $ 86.7\%$ & $33.1\%$ & $25.6\%$ & $ 19.3\%$\\
		simex - 1 ($v=4.8$) & $93.1\%$ & $90.3\%$ & $ 86.6\%$ & $33.3\%$ & $25.7\%$ & $ 19.3\%$\\
		simex - 1 ($v=8$) & $93.2\%$ & $90.3\%$ & $ 86.4\%$ & $33.8\%$ & $25.9\%$ & $ 19.3\%$\\
		naive - 2 & $93.4\%$ & $90.9\%$ & $ 87.6\%$ & $35.9\%$ & $28.2\%$ & $ 21.7\%$\\
		simex - 2 ($v=4.8$) & $93.2\%$ & $90.7\%$ & $ 87.4\%$ & $34.6\%$ & $27.4\%$ & $ 21.2\%$\\
		simex - 2 ($v=8$) & $93.2\%$ & $90.7\%$ & $ 87.3\%$ & $34.9\%$ & $27.6\%$ & $ 21.3\%$\\
	\end{tabular}}
	\caption{Estimated cure probability for given PSA level and stage. The  naive and  simex estimators are computed using the maximum likelihood (1) or the  presmoothing (2) approach.}
	\label{tab:cure_prostate}
\end{table}
\section*{Acknowledgements}
The authors acknowledge  financial support from the European Research Council (2016–2021, Horizon 2020, grant agreement 694409). For the simulations we used the infrastructure of the Flemish
Supercomputer Center (VSC).

\section*{Supplemetary Material}
Supporting information may be found in the online appendix.
This document contains the proofs of the theorems in Section~\ref{sec:asymptotics} and  additional simulation results.

\bibliographystyle{biometrika}
\bibliography{cure_models}

\clearpage
\setcounter{equation}{0}
\renewcommand{\theequation}{S\arabic{equation}} \setcounter{table}{0}
\renewcommand{\thetable}{S\arabic{table}}  
\setcounter{page}{1}
\newpage

\thispagestyle{plain}
\centerline{ }
\centerline{ }

\centerline{
	\LARGE\bf A simulation-extrapolation approach }
\smallskip
\centerline{\LARGE\bf for the mixture cure model with mismeasured covariates}
\bigskip
\centerline{\LARGE Supplementary Material}
\bigskip
\centerline{\Large Eni Musta and Ingrid Van Keilegom}
\medskip
\centerline{\it ORSTAT, KU Leuven }
\bigskip

\appendix

This supplement is organized as follows.
Appendix A contains proofs of Theorems~\ref{theo:consistency}-\ref{theo:bb}.
Appendix B collects additional simulation results, that were omitted from the main paper due to page limits.

\section{Proofs }
\label{sec:proofs}
\textit{Proof of Theorem \ref{theo:consistency}.} 
For a fixed $\lambda$ and $b$, from condition (A1) we have
\[
\Vert\hat\gamma_{\lambda,b}-\gamma_\lambda\Vert\to0,\qquad \Vert\hat\beta_{\lambda,b}-\beta_\lambda\Vert\to 0\qquad\text{ and} \qquad\sup_{t\in[0,\tau]}|\hat\Lambda_{\lambda,b}(t)-\Lambda_\lambda(t)|\to 0
\]
with probability one. By definition of $\hat\gamma_{\lambda}$, $\hat\beta_{\lambda}$ and $\hat\Lambda_{\lambda}$ as averages over the correspondent values for $b=1,\dots,B$ (see \eqref{eqn:averages}) and Slutsky's theorem, it follows that for each $\lambda\in\{\lambda_1,\dots,\lambda_K\}$
\[
\Vert\hat\gamma_{\lambda}-\gamma_\lambda\Vert\to0,\qquad \Vert\hat\beta_{\lambda}-\beta_\lambda\Vert\to 0\qquad\text{ and} \qquad\sup_{t\in[0,\tau]}|\hat\Lambda_{\lambda}(t)-\Lambda_\lambda(t)|\to 0
\]
almost surely. Next we first focus on consistency of $\hat\gamma_{\mathrm{simex}}$. Since we are assuming that $g_{\gamma}(a_\gamma^*,\lambda)=(g_{\gamma,1}(a_{\gamma_1}^*,\lambda),\dots,g_{\gamma,p}(a_{\gamma_p}^*,\lambda))^T$ is the true extrapolation function, we have $\gamma_\lambda=g_{\gamma}(a_\gamma^*,\lambda)$ and  $\gamma_0=g_\gamma(a_\gamma^*,-1)$. On the other hand, $\hat\gamma_{\mathrm{simex}}=g_{\gamma}(\hat{a}_\gamma,-1)$, where $\hat{a}_\gamma$ is the least squares  estimator of $a_\gamma^*$, i.e. it solves
\[
\Psi_n(a_\gamma)=\dot{g}_{\gamma}(a_{\gamma},{\bm\lambda})^T\left\{g_{\gamma}(a_{\gamma},\bm{\lambda})-\hat\gamma_{\bm{\lambda}}\right\}=0
\]
where $\hat\gamma_{\bm{\lambda}}=(\hat\gamma^T_{\lambda_1},\dots,\hat\gamma^T_{\lambda_K})^T$, $g_{\gamma}(a_{\gamma},\bm{\lambda})=(g_{\gamma}(a_{\gamma},\lambda_1)^T,\dots,g_{\gamma}(a_{\gamma},\lambda_K)^T)^T$ and $\dot{g}_{\gamma}(a_{\gamma},{\bm\lambda})$ is the $pK\times p\,\mathrm{dim}(a_\gamma)$ matrix of partial derivatives of the elements of $g_{\gamma}(a_{\gamma},\bm{\lambda})$ with respect to the elements of $a_\gamma$. Moreover, the true parameters $a_\gamma^*$ are the solution of
\[
\Psi(a_\gamma)=\dot{g}_{\gamma}(a_{\gamma},{\bm\lambda})^T\left\{g_{\gamma}(a_{\gamma},\bm{\lambda})-\gamma_{\bm{\lambda}}\right\}=0
\]
and 
\[
\sup_{a_\gamma}\Vert\Psi_n(a)-\Psi(a)\Vert\leq \sup_{a_\gamma}\Vert\dot{g}_{\gamma}(a_{\gamma},{\bm\lambda})\Vert \Vert\hat\gamma_{\bm{\lambda}}-\gamma_{\bm{\lambda}}\Vert\to 0 \qquad\text{a.s.}
\]
Hence, if $a^*_\gamma$ is the unique solution of $\Psi(a_\gamma)=0$, it follows that $\hat{a}_\gamma\to a^*_\gamma$ with probability one. From the continuous mapping theorem it follows that 
\[
\Vert \hat{\gamma}_{\mathrm{simex}}-\gamma_0\Vert\to 0\qquad\text{a.s.}
\]
Consistency of $\hat\beta_{\mathrm{simex}}$ can be proven in the same way. For the function $\hat\Lambda_{\mathrm{simex}}$ we suppose that for every $t\in[0,\tau]$, $\Lambda_\lambda(t)$ can be specified by a function $g_{\Lambda,t}(a_t,\lambda)$ depending on a parametric vector $a_t$ and $\Lambda_0(t)=g_{\Lambda,t}(a_t,-1)$. Hence, arguing as above, for any fixed $t\in[0,\tau_0]$ we can show that 
\[
|\hat{\Lambda}_{\mathrm{simex}}(t)-\Lambda_0(t)|\to 0\qquad\text{a.s.}
\] 
Uniform consistency on the compact $[0,\tau]$ follows from the fact that $\Lambda_0$ is continuous and $\Lambda_{\mathrm{SIMEX}}$ is non-decreasing. \hfill$\square$

\textit{Proof of Theorem \ref{theo:normality}.}
For fixed $\lambda$ and $b$, from condition (C2) we have
\[
\begin{split}
&h^T_2(\hat{\gamma}_{\lambda,b}-\gamma_\lambda)+h_3^T(\hat{\beta}_{\lambda,b}-\beta_\lambda)+\int_0^\tau h_1(s)\dd(\hat\Lambda_{\lambda,b}-\Lambda_\lambda)(s)\\
&=\frac{1}{n}\sum_{i=1}^n\Psi(Y_i,\Delta_i,W_{i,\lambda,b},h_1,h_2,h_3)+o_P(n^{-1/2})
\end{split}
\]	
uniformly over $(h_1,h_2,h_3)\in\mathcal{H}_{\mathfrak{m}}$. As a result, 
\[
\begin{split}
&h^T_2(\hat{\gamma}_{\lambda}-\gamma_\lambda)+h_3^T(\hat{\beta}_{\lambda}-\beta_\lambda)+\int_0^\tau h_1(s)\dd(\hat\Lambda_{\lambda}-\Lambda_\lambda)(s)\\
&= h^T_2\left(\frac{1}{B}\sum_{b=1}^B\hat{\gamma}_{\lambda,b}-\gamma_\lambda\right)+h_3^T\left(\frac{1}{B}\sum_{b=1}^B\hat{\beta}_{\lambda,b}-\beta_\lambda\right)+\int_0^\tau h_1(s)\dd\left(\frac{1}{B}\sum_{b=1}^B\hat\Lambda_{\lambda,b}-\Lambda_\lambda\right)(s)\\
&=\frac{1}{n}\sum_{i=1}^n\left\{\frac{1}{B}\sum_{b=1}^B\Psi(Y_i,\Delta_i,W_{i,\lambda,b},h_1,h_2,h_3)\right\}+o_P(n^{-1/2}).
\end{split}
\]
Since sum of Donsker classes is Donsker (see Lemma 2.10.6 in \cite{VW96}), it follows that the process 
\[
n^{1/2}\left\{h^T_2(\hat{\gamma}_{\lambda}-\gamma_\lambda)+h_3^T(\hat{\beta}_{\lambda}-\beta_\lambda)+\int_0^\tau h_1(s)\dd(\hat\Lambda_{\lambda}-\Lambda_\lambda)(s)\right\}
\]
converges weakly to a zero-mean Gaussian process $G_\lambda$ indexed by $h=(h_1,h_2,h_3)\in\mathcal{H}_{\mathfrak{m}}$ and covariance function
\[
\begin{split}
&Cov\left(G_\lambda(h_1,h_2,h_3),G_\lambda(h_1^*,h_2^*,h_3^*)\right)\\
&=\E\left[\left\{\frac{1}{B}\sum_{b=1}^B\Psi(Y,\Delta,W_{\lambda,b},h_1,h_2,h_3)\right\}\left\{\frac{1}{B}\sum_{b=1}^B\Psi(Y,\Delta,W_{\lambda,b},h_1^*,h_2^*,h_3^*)\right\}\right].
\end{split}
\]
Moreover, the $K$ dimensional vector 
\[
n^{1/2}\begin{Bmatrix}
h^T_2(\hat{\gamma}_{\lambda_1}-\gamma_{\lambda_1})+h_3^T(\hat{\beta}_{\lambda_1}-\beta_{\lambda_1})+\int_0^\tau h_1(s)\dd(\hat\Lambda_{\lambda_1}-\Lambda_{\lambda_1})(s)\\
\vdots\\
h^T_2(\hat{\gamma}_{\lambda_K}-\gamma_{\lambda_K})+h_3^T(\hat{\beta}_{\lambda_K}-\beta_{\lambda_K})+\int_0^\tau h_1(s)\dd(\hat\Lambda_{\lambda_K}-\Lambda_{\lambda_K})(s)
\end{Bmatrix}
\]
converges to a $K$ dimensional Gaussian process $G_{\bm{\lambda}}$ with mean zero and covariance function between the $i$th and the $jth$ component 
	\[
\E\left[\left\{\frac{1}{B}\sum_{b=1}^B\Psi(Y,\Delta,W_{\lambda_i,b},h_1,h_2,h_3)\right\}\left\{\frac{1}{B}\sum_{b=1}^B\Psi(Y,\Delta,W_{\lambda_j,b},h_1^*,h_2^*,h_3^*)\right\}\right].
	\]
	In particular, if we take $h_1\equiv 0$, $h_3=0$ and $h_2=(0,\dots,0,1,0,\dots,0)$ with $h_2$ containing 1 at the $j$th position ($j=1,\dots,p$) and 0 elsewhere, we obtain the weak convergence of $n^{1/2}(\hat\gamma_{\bm{\lambda}}-\gamma_{\bm{\lambda}})$ to a multivariate normal random variable with mean zero and covariance matrix $\Sigma_{\gamma,\bm{\lambda}}$. With the same reasoning we also obtain $n^{1/2}(\hat\beta_{\bm{\lambda}}-\beta_{\bm{\lambda}})\xrightarrow{d}N(0,\Sigma_{\beta,\bm{\lambda}})$. For $\hat\Lambda_{\bm{\lambda}}$ we consider the class 
		\[
		\left\{(h_1,h_2,h_3)\in\mathcal{H}_{\mathfrak{m}}\,:\, h_2=h_3=0 \,\text{ and } \,h_1(s)=\1_{\{s\leq t\}},\, t\in[0,\tau]\right\}
		\]
		and obtain the weak convergence of $n^{1/2}\{\hat{\Lambda}_{\bm{\lambda}}(t)-\Lambda_{\bm{\lambda}}(t)\}$ to a Gaussian process $\mathcal{G}_{\bm{\lambda}}$ indexed by $t\in[0,\tau]$.
		
		Next we prove the asymptotic normality of $\hat\gamma_{\mathrm{simex}}$. Since we are assuming that $g_{\gamma}(a_\gamma^*,\lambda)=(g_{\gamma,1}(a_{\gamma_1}^*,\lambda),\dots,g_{\gamma,p}(a_{\gamma_p}^*,\lambda))^T$ is the true extrapolation function, we have $\gamma_\lambda=g_{\gamma}(a_\gamma^*,\lambda)$ and  $\gamma_0=g_\gamma(a_\gamma^*,-1)$. On the other hand, $\hat\gamma_{\mathrm{simex}}=g_{\gamma}(\hat{a}_\gamma,-1)$, where $\hat{a}_\gamma$ is the least squares  estimator of $a_\gamma^*$, i.e. it solves
		\begin{equation}
		\label{eqn:psi_n=0}
		\Psi_n(a_\gamma)=\dot{g}_{\gamma}(a_{\gamma},{\bm\lambda})^T\left\{g_{\gamma}(a_{\gamma},\bm{\lambda})-\hat\gamma_{\bm{\lambda}}\right\}=0
		\end{equation}
		where $\hat\gamma_{\bm{\lambda}}=(\hat\gamma^T_{\lambda_1},\dots,\hat\gamma^T_{\lambda_K})^T$, $g_{\gamma}(a_{\gamma},\bm{\lambda})=(g_{\gamma}(a_{\gamma},\lambda_1)^T,\dots,g_{\gamma}(a_{\gamma},\lambda_K)^T)^T$ and $\dot{g}_{\gamma}(a_{\gamma},{\bm\lambda})$ is the $pK\times p\,\mathrm{dim}(a_\gamma)$ matrix of partial derivatives of the elements of $g_{\gamma}(a_{\gamma},\bm{\lambda})$ with respect to the elements of $a_\gamma$. Since $\hat{a}_\gamma$ solves equation \eqref{eqn:psi_n=0} and $\hat{a}_\gamma\to a_\gamma^*$ with probability one (see proof of Theorem \ref{theo:consistency}), if $\dot{g}_\gamma(a_\gamma,\bm{\lambda})$ is bounded and continuous w.r.t. $a_\gamma$ and $\dot{g}_\gamma(a_\gamma,\bm{\lambda})^T\dot{g}_\gamma(a_\gamma,\bm{\lambda})$ is invertible, we have
		\[
		n^{1/2}(\hat{a}_\gamma- a_\gamma^*)=\left\{\dot{g}_\gamma(a_\gamma^*,\bm{\lambda})^T\dot{g}_\gamma(a_\gamma^*,\bm{\lambda})\right\}^{-1}\dot{g}_\gamma(a_\gamma^*,\bm{\lambda})^Tn^{1/2}(\hat\gamma_{\bm{\lambda}}-\gamma_{\bm{\lambda}})+o_P(1).
		\]
		As a result,
		\[
		n^{1/2}(\hat{a}_\gamma- a_\gamma^*)\xrightarrow{d}\left\{\dot{g}_\gamma(a_\gamma^*,\bm{\lambda})^T\dot{g}_\gamma(a_\gamma^*,\bm{\lambda})\right\}^{-1}\dot{g}_\gamma(a_\gamma^*,\bm{\lambda})^TN(0,\Sigma_{\gamma,\bm{\lambda}}).
		\]
	Finally, using the delta method, we obtain
	\[
	n^{1/2}(\hat\gamma_{\mathrm{simex}}-\gamma_0)\xrightarrow{d}\dot{g}_\gamma(a_\gamma^*,-1)\left\{\dot{g}_\gamma(a_\gamma^*,\bm{\lambda})^T\dot{g}_\gamma(a_\gamma^*,\bm{\lambda})\right\}^{-1}\dot{g}_\gamma(a_\gamma^*,\bm{\lambda})^TN(0,\Sigma_{\gamma,\bm{\lambda}}),
	\]
	meaning that $n^{1/2}(\hat\gamma_{\mathrm{simex}}-\gamma_0)$ converges weakly to a multivariate normal random variable with mean zero and covariance matrix
	\begin{equation}
	\label{eqn:sigma_gamma}
	\begin{split}
	\Sigma_\gamma&=\dot{g}_\gamma(a_\gamma^*,-1)\left\{\dot{g}_\gamma(a_\gamma^*,\bm{\lambda})^T\dot{g}_\gamma(a_\gamma^*,\bm{\lambda})\right\}^{-1}\dot{g}_\gamma(a_\gamma^*,\bm{\lambda})^T\\
	&\qquad\qquad\qquad\,\times\Sigma_{\gamma,\bm{\lambda}}\,\dot{g}_\gamma(a_\gamma^*,\bm{\lambda})\left\{\dot{g}_\gamma(a_\gamma^*,\bm{\lambda})^T\dot{g}_\gamma(a_\gamma^*,\bm{\lambda})\right\}^{-1}\dot{g}_\gamma(a_\gamma^*,-1)^T.
	\end{split}
	\end{equation}
	In the same way it can be shown that 
	$n^{1/2}(\hat\beta_{\mathrm{simex}}-\beta_0)$ converges weakly to a multivariate normal random variable with mean zero and covariance matrix
	\begin{equation}
	\label{eqn:sigma_beta}
	\begin{split}
	\Sigma_\beta&=\dot{g}_\beta(a_\beta^*,-1)\left\{\dot{g}_\beta(a_\beta^*,\bm{\lambda})^T\dot{g}_\beta(a_\beta^*,\bm{\lambda})\right\}^{-1}\dot{g}_\beta(a_\beta^*,\bm{\lambda})^T\\
	&\qquad\qquad\qquad\,\times\Sigma_{\beta,\bm{\lambda}}\,\dot{g}_\beta(a_\beta^*,\bm{\lambda})\left\{\dot{g}_\beta(a_\beta^*,\bm{\lambda})^T\dot{g}_\beta(a_\beta^*,\bm{\lambda})\right\}^{-1}\dot{g}_\beta(a_\beta^*,-1)^T.
	\end{split}
	\end{equation}
	Similarly, for the nonparametric component we have 
		\[
	n^{1/2}(\hat{a}_t- a_t^*)=\left\{\dot{g}_{\Lambda,t}(a_t^*,\bm{\lambda})^T\dot{g}_{\Lambda,t}(a_t^*,\bm{\lambda})\right\}^{-1}\dot{g}_{\Lambda,t}(a_t^*,\bm{\lambda})^Tn^{1/2}(\hat\Lambda_{\bm{\lambda}}(t)-\Lambda_{\bm{\lambda}}(t))+o_P(1)
	\]
for all $t\in[0,\tau]$. From the weak convergence of the process $n^{1/2}(\hat\Lambda_{\bm{\lambda}}-\Lambda_{\bm{\lambda}})$, it follows that $n^{1/2}(\hat{a}_t- a_t^*)$ converges in distribution to the Gaussian process 
\[
\left\{\dot{g}_{\Lambda,t}(a_t^*,\bm{\lambda})^T\dot{g}_{\Lambda,t}(a_t^*,\bm{\lambda})\right\}^{-1}\dot{g}_{\Lambda,t}(a_t^*,\bm{\lambda})^T\mathcal{G}_{\bm{\lambda}}.
\]
Once more, the delta method yields that $n^{1/2}(\hat\Lambda_{\mathrm{simex}}-\Lambda_{0})$ converges weakly to the Gaussian process
\begin{equation}
\label{eqn:limit_G}
\mathcal{G}=\dot{g}_{\Lambda,t}(a_t^*,-1)\left\{\dot{g}_{\Lambda,t}(a_t^*,\bm{\lambda})^T\dot{g}_{\Lambda,t}(a_t^*,\bm{\lambda})\right\}^{-1}\dot{g}_{\Lambda,t}(a_t^*,\bm{\lambda})^T\mathcal{G}_{\bm{\lambda}}.
\end{equation}
	\hfill$\square$
	
\textit{Proof of Theorem \ref{theo:smcure}.}
	Let $\Upsilon_0=(\gamma_0,\beta_0,\Lambda_0)$, $\theta_0=(\gamma_0,\beta_0)$, $\hat\theta_n=(\hat\gamma_n,\hat\beta_n)$ and $\mathcal{H}_{\mathfrak{m}}$ as in (A2). Define the continuous linear operator $\sigma=(\sigma_1,\sigma_2)$ from $\mathcal{H}_{\mathfrak{m}}$ to $\mathcal{H}_{\mathfrak{m}}$ of the form 
		\begin{equation}
		\label{eqn:sigma_1}
	\begin{split}
	\sigma_1(h)(t)&=\E\left[\1_{\{Y\geq t\}}V(t,\Upsilon_0)(h)g(t,\Upsilon_0)e^{\beta^T_0Z}\right]\\
	&\quad-\E\left[\int_t^{\tau}\1_{\{Y\geq s\}}V(t,\Upsilon_0)(h)g(s,\Upsilon_0)\{1-g(s,\Upsilon_0)\}e^{2\beta^T_0Z}\dd\Lambda_0(s)\right]
	\end{split}
	\end{equation}
	and
	\begin{equation}
	\label{eqn:sigma_2}
	\sigma_2(h)(t)=\E\left[\int_0^{\tau}\1_{\{Y\geq t\}}W(t,\Upsilon_0)V(t,\Upsilon_0)(h)g(t,\Upsilon_0)e^{\beta^T_0Z}\dd\Lambda_0(t)\right]
	\end{equation}
	where
	\begin{equation}
	\label{def:g}
	g(t,\Lambda,\beta,\gamma)=\frac{\phi(\gamma,X)\exp\left(-\Lambda(t) \exp\left(\beta^T Z\right)\right)}{1-\phi(\gamma,X)+\phi(\gamma,X)\exp\left(-\Lambda(t) \exp\left(\beta^T Z\right)\right)},
	\end{equation}
	\[
	V(t,\Upsilon_0)(h)=h_1(t)-\left\{1-g(t,\Upsilon_0)\right\}e^{\beta^T_0Z}\int_0^th_1(s)\dd\Lambda_0(s)+(h^T_2,h^T_3)W(t,\Upsilon_0)
	\]
	and
	\[
	W(t,\Upsilon_0)=\left(\left\{1-g(t,\Upsilon_0)\right\}X^T,\left[1-\left\{1-g(t,\Upsilon_0)\right\}e^{\beta^T_0Z}\Lambda_0(t)\right]Z^T\right)^T.
	\]
	Note that in our case $X=(W_\lambda^{(1)},W_\lambda^{(2)})$ and $Z=(W_\lambda^{(1)},W_\lambda^{(2)})$. 
	In the proof of Theorem 2 in \cite{Lu2008} (page 572) it is shown that 
	\begin{equation}
	\label{eqn:main_Lu}
	\int_0^{\tau}\sigma_1(h)(t)\,\dd\sqrt{n}(\Lambda_n-\Lambda_0)(t)+\sqrt{n}(\hat\theta_n-\theta_0)^T\sigma_2(h)=\sqrt{n}\left\{S_n(\Upsilon_0)-S(\Upsilon_0)\right\}(h)+o_P(1),
	\end{equation}
	where
	\[
	\sqrt{n}\left\{S_n(\Upsilon_0)-S(\Upsilon_0)\right\}(h_1,h_2,h_3)=\int f_h(y,\delta,x,z)\,\dd\sqrt{n}(\p_n-\p)(y,\delta,x,z)
	\]
	and $\{f_h(y,\delta,x,z), h\in\mathcal{H}_{\mathfrak{m}}\}$ is a uniformly bounded Donsker class such that 
	\[
	\mathrm{E}\left[f_h(Y,\Delta,X,Z)\right]=S(\Upsilon_0)=0.
	\] 
	In \cite{Lu2008} it is also shown that $\sigma$ is invertible with inverse $ \sigma^{-1}=(\sigma_1^{-1},\sigma_2^{-1})$.  Hence, for all $h\in\mathcal{H}_{\mathfrak{m}}$, if in \eqref{eqn:main_Lu} we replace $h$ by $\sigma^{-1}(h)$, we obtain
	\[
	\begin{split}
	&\int_0^{\tau}h_1(t)\dd(\Lambda_n(t)-\Lambda_0(t))+h^T_2(\hat\gamma_n-\gamma_0)+h^T_3(\hat\beta_n-\beta_0)\\
	&=\int f_{\sigma^{-1}(h)}(y,\delta,x,z)\,\dd(\p_n-\p)(y,\delta,x,z)+o_P(n^{-1/2})
	\end{split}
	\]
	and (A2) holds with 
	\begin{equation}
	\label{eqn:psi_smcure}
	\begin{split}
	\Psi_\lambda(y,\delta,w,h_1,h_2,h_3)&=f_{\sigma^{-1}(h)}\left(y,\delta,(w^{(1)},w^{(2)}),(w^{(2)},w^{(3)})\right)
	\end{split}
	\end{equation}
	\hfill$\square$

\textit{Proof of Theorem \ref{theo:bb}.}
	In \cite{MPK20} it is shown that 
	\begin{equation}
	\label{eqn:gama_bb}
	\hat\gamma_n-\gamma_0=-(\Gamma_1^T\Gamma_1)^{-1}\Gamma^T_1	\int \psi(y,\delta,x)\,\d(\p_n-\p)(y,\delta,x,z)+ o_P(n^{-1/2})
	\end{equation}
	(see their equation {(A.33)}), where
	\begin{equation*}
	\label{def:gradient_m}
	\Gamma_1=-\mathrm{E}\left[\left\{\frac{1}{\phi(\gamma_0,X)}+\frac{1}{1-\phi(\gamma_0,X)}\right\}\nabla_\gamma\phi(\gamma_0,X)\nabla_\gamma\phi(\gamma_0,X)^T\right],
	\end{equation*}
	\[
	\psi(y,\delta,x)=-\left\{\frac{\Delta\1_{\{y\leq\tau_0\}}}{{1-H(y|x)}}-\int_0^{y\wedge \tau_0}\frac{H_1(ds|x)}{{(1-H(s|x))^2}}\right\}\frac{1}{\phi(\gamma_0,x)}\nabla_\gamma\phi(\gamma_0,x)
	\]
	with   {$H_k(t|x)=\p\left(Y\leq t,\Delta=k|X=x\right)$}  for $k=0,1$ and  {$H(t|x)=H_1(t|x)+H_0(t|x)$}. Moreover we have $\mathrm{E}\left[\psi(Y,\Delta,X)\right]=0$. 
	
		Let $\Upsilon_0=(\gamma_0,\beta_0,\Lambda_0)$ and 
		$\mathcal{\tilde{H}}_{\mathfrak{m}}=\{\tilde{h}=(h_1,h_3)\in BV[0,\tau_0]\times\R^q\,:\, \Vert h_1\Vert_v+\Vert h_3\Vert_{L_1}\leq \mathfrak{m}\}$.  Define the continuous linear operator $\sigma=(\sigma_1,\sigma_2)$ from $\mathcal{\tilde{H}}_{\mathfrak{m}}$ to $\mathcal{\tilde{H}}_{\mathfrak{m}}$ as in \eqref{eqn:sigma_1}, \eqref{eqn:sigma_2} with 
	\[
	V(t,\Upsilon_0)(h)=h_1(t)-\left\{1-g(t,\Upsilon_0)\right\}e^{\beta^T_0Z}\int_0^th_1(s)\dd\Lambda_0(s)+h^T_3W(t,\Upsilon_0)
	\]
	and
	\[
	W(t,\Upsilon_0)=\left[1-\left\{1-g(t,\Upsilon_0)\right\}e^{\beta^T_0Z}\Lambda_0(t)\right]Z.
	\]
	From  equations {(A37)-(A38)} in \cite{MPK20} we have 
	\begin{equation*}
	\label{eqn:main_bb}
	\int_0^{\tau}\sigma_1(\tilde{h})(t)\,\dd\sqrt{n}(\Lambda_n-\Lambda_0)(t)+\sqrt{n}(\hat\beta_n-\beta_0)^T\sigma_2(\tilde{h})=\sqrt{n}\left\{\hat{S}_n(\Upsilon_0)-S(\Upsilon_0)\right\}(\tilde{h})+o_P(1),
	\end{equation*}
	where
	\[
	\sqrt{n}\left\{\hat{S}_n(\Upsilon_0)-S(\Upsilon_0)\right\}(h_1,h_3)=\int \tilde{f}_{\tilde{h}}(y,\delta,x,z)\,\dd\sqrt{n}(\p_n-\p)(y,\delta,x,z)
	\]
	for some uniformly bounded Donsker class $\{f_{\tilde{h}}(y,\delta,x,z), \tilde{h}\in\mathcal{\tilde{H}}_{\mathfrak{m}}\}$ with $\mathrm{E}[\tilde{f}_{\tilde{h}}(Y,\Delta,X,Z)]=0$.  Hence, if  we replace $\tilde{h}$ by $\sigma^{-1}(\tilde{h})$, we obtain
	\[
	\int_0^{\tau}h_1(t)\dd(\Lambda_n(t)-\Lambda_0(t))+h^T_3(\hat\beta_n-\beta_0)=\int \tilde{f}_{\sigma^{-1}(\tilde{h})}(y,\delta,x,z)\,\dd(\p_n-\p)(y,\delta,x,z)+o_P(n^{-1/2}).
	\]
	Note that in our case $x=(w_\lambda^{(1)},w_\lambda^{(2)})$ and $z=(w_\lambda^{(1)},w_\lambda^{(2)})$. Moreover, if $h=(h_1,h_2,h_3)\in\mathcal{H}_{\mathfrak{m}}$, then $\tilde{h}=(h_1,h_3)\in\mathcal{\tilde{H}}_{\mathfrak{m}}$. 
	It follows that (A2) holds with 
	\begin{equation}
	\label{eqn:psi_bb}
	\begin{split}
	\Psi_\lambda(y,\delta,w,h_1,h_2,h_3)&=-h^T_2(\Gamma_1^T\Gamma_1)^{-1}\Gamma^T_1	 \psi\left(y,\delta,(w^{(1)},w^{(2)})\right)\\
	&\quad +\tilde{f}_{\sigma^{-1}((h_1,h_3))}\left(y,\delta,(w^{(1)},w^{(2)}),(w^{(2)},w^{(3)})\right)
	\end{split}
	\end{equation}
		\hfill$\square$
\section{Additional simulation results}
In this section we report the simulation results for sample size  $n=400$ in Model~1, and results for Models 2-5 ($n=200$) that were omitted from the main paper.
\begin{table}[h]
	\centering
	\scalebox{0.85}{
		\begin{tabular}{ccrrrrrrrrrrrr}
			&  & & & & & & & && &&&\\[-8pt]
			&&\multicolumn{3}{c}{naive - 1}&\multicolumn{3}{c}{naive - 2}&\multicolumn{3}{c}{simex - 1}&\multicolumn{3}{c}{simex - 2}\\
			& Par. &  Bias & Var. & MSE & Bias & Var. & MSE & Bias & Var. & MSE&Bias & Var. & MSE\\[2pt]
			& & & & & & & & & && &&\\[-8pt]
			$1/1/1$  & $\gamma_1 $ & $0.8 $  & $1.9$ & $ 1.9$ & $ 0.5 $ & $1.8 $ &  $  1.8 $ & $0.9  $ & $  1.9 $ & $ 1.9  $ &$ 0.7$&$ 1.9$&$1.9 $\\
			& $\gamma_2 $ & $ -4.1  $  & $ 1.4  $ & $ 1.5  $ & $-3.7 $  & $ 1.3  $ & $ 1.4 $ & $ -1.8 $ &$2.5 $&$ 2.5$&$-1.7 $&$2.5 $&$ 2.5$\\
			& $\beta $ & $  -44.3 $ & $0.3 $ & $ 19.9 $ & $-44.3   $ & $0.3    $ & $20.0 $ & $-20.1  $ & $ 0.9  $ & $4.9  $ & $ -20.2  $& $ 0.9 $ &$ 4.9$\\
			& & & & & & & & & && &&\\[-8pt]
			$1/1/2$  & $\gamma_1 $ & $1.8$  & $2.7$ & $ 2.7$ & $0.5  $ & $2.6 $ &  $ 2.6  $ & $ 1.5 $ & $ 2.6  $ & $ 2.6  $ &$0.7 $&$3.0 $&$3.0 $\\
			& $\gamma_2 $ & $  -4.1 $  & $ 2.0  $ & $ 2.1  $ & $-1.8 $  & $ 2.0  $ & $ 2.0 $ & $ -1.3 $ &$3.6 $&$3.6 $&$1.0 $&$ 4.3$&$4.3 $\\
			& $\beta $ & $  -43.2 $ & $0.4 $ & $ 19.1 $ & $  -43.5 $ & $  0.4  $ & $19.3 $ & $ -19.2 $ & $ 1.1  $ & $ 4.8 $ & $ -19.4  $& $ 1.1 $ &$4.9 $\\
			& & & & & & & & & && &&\\[-8pt]
			$1/2/1$  & $\gamma_1 $ & $-0.3 $  & $1.3$ & $1.3$ & $ -0.6 $ & $1.3 $ &  $  1.3 $ & $ -0.4 $ & $  1.3 $ & $  1.3 $ &$ -0.5$&$1.3 $&$ 1.3$\\
			& $\gamma_2 $ & $ -3.7  $  & $  0.8 $ & $ 0.9  $ & $ -3.8$  & $  0.7 $ & $0.9  $ & $-1.5  $ &$1.5 $&$ 1.5$&$ -1.8$&$1.4 $&$1.4 $\\
			& $\beta $ & $  -43.9 $ & $0.6 $ & $ 19.9 $ & $ -43.9  $ & $ 0.4   $ & $19.3 $ & $ -19.8 $ & $ 1.6  $ & $ 5.5 $ & $ -19.8  $& $ 1.5 $ &$ 5.5$\\
			& & & & & & & & & && &&\\[-8pt]
			$1/2/2$  & $\gamma_1 $ & $-0.2$  & $2.1$ & $ 2.1$ & $ -1.4 $ & $2.1 $ &  $ 2.1  $ & $ -0.5 $ & $2.1   $ & $ 2.1  $ &$ -1.4$&$ 2.2$&$ 2.3$\\
			& $\gamma_2 $ & $ -3.1  $  & $  1.3 $ & $1.4   $ & $-2.3 $  & $  1.2 $ & $ 1.3 $ & $  -0.5$ &$2.4 $&$ 2.4$&$0.3 $&$ 2.5$&$2.5 $\\
			& $\beta $ & $ -42.5  $ & $ 0.9$ & $ 19.0 $ & $ -42.8  $ & $ 0.9   $ & $ 19.3$ & $-18.6  $ & $ 2.4  $ & $ 5.9 $ & $ -18.9  $& $ 2.4 $ &$ 5.9$\\
			& & & & & & & & & && &&\\[-8pt]
			$2/1/1$  & $\gamma_1 $ & $-2.0 $  & $1.8$ & $ 1.9$ & $-2.3  $ & $ 1.8$ &  $ 1.8  $ & $ -0.4 $ & $  1.9 $ & $  1.9 $ &$-0.8 $&$1.9 $&$1.9 $\\
			& $\gamma_2 $ & $ -17.3  $  & $ 1.2  $ & $   4.2$ & $ -17.9$  & $   1.2$ & $ 4.4 $ & $ -5.7 $ &$2.3 $&$2.7 $&$-6.8 $&$2.4 $&$2.9 $\\
			& $\beta $ & $ -45.0  $ & $0.3 $ & $20.6  $ & $ -45.0  $ & $ 0.3   $ & $ 20.6$ & $-20.8  $ & $0.9   $ & $5.2  $ & $-20.8   $& $ 0.9 $ &$5.2 $\\
			& & & & & & & & & && &&\\[-8pt]
			$2/1/2$  & $\gamma_1 $ & $-1.2$  & $2.3$ & $ 2.3$ & $-2.3  $ & $ 2.3$ &  $ 2.3  $ & $0.3  $ & $  2.4 $ & $  2.4 $ &$ -0.7$&$ 2.5$&$2.5 $\\
			& $\gamma_2 $ & $ -17.2  $  & $ 1.7  $ & $ 4.7  $ & $ -16.7$  & $ 1.6  $ & $4.4  $ & $ -5.2 $ &$ 3.3$&$ 3.6$&$ -5.3$&$3.5 $&$ 3.8$\\
			& $\beta $ & $ -44.2  $ & $ 0.4$ & $ 19.9 $ & $ -44.3  $ & $  0.4  $ & $ 20.0$ & $ -20.1 $ & $ 1.1  $ & $5.1  $ & $-20.1   $& $  1.1$ &$5.1 $\\
			& & & & & & & & & && &&\\[-8pt]
			$2/2/1$  & $\gamma_1 $ & $-0.5 $  & $1.3$ & $ 1.3$ & $-0.8  $ & $1.3 $ &  $ 1.3  $ & $ -0.5 $ & $ 1.3  $ & $ 1.3  $ &$-0.7 $&$ 1.4$&$ 1.4$\\
			& $\gamma_2 $ & $  -17.3 $  & $ 0.9  $ & $   3.9$ & $-18.4 $  & $ 0.9  $ & $4.2  $ & $-5.6  $ &$1.8 $&$2.1 $&$-7.5 $&$1.7 $&$ 2.3$\\
			& $\beta $ & $ -44.5  $ & $ 0.6$ & $20.4  $ & $  -44.4 $ & $  0.6  $ & $20.4 $ & $ -20.0 $ & $ 1.7  $ & $ 5.7 $ & $ -19.9  $& $ 1.7 $ &$5.6 $\\
			& & & & & & & & & && &&\\[-8pt]
			$2/2/2$  & $\gamma_1 $ & $0.0 $  & $2.4$ & $ 2.4$ & $ -1.5 $ & $ 2.3$ &  $  2.4 $ & $-0.2  $ & $2.4   $ & $ 2.4  $ &$-1.5 $&$2.6 $&$2.7 $\\
			& $\gamma_2 $ & $ -17.1  $  & $  1.5 $ & $  4.4 $ & $ -17.6$  & $  1.6 $ & $ 4.7 $ & $-5.1  $ &$3.0 $&$ 3.3$&$-5.9 $&$ 3.5$&$ 3.9$\\
			& $\beta $ & $  -42.8 $ & $ 1.0$ & $19.3  $ & $ -42.9  $ & $  1.0  $ & $ 19.5$ & $ -18.4 $ & $ 2.6  $ & $6.0  $ & $ -18.6  $& $ 2.6 $ &$6.1 $\\
			& & & & & & & & & && &&\\[-8pt]
			$3/1/1$  & $\gamma_1 $ & $-37.0$  & $3.4$ & $17.1$ & $ -38.7 $ & $ 3.2$ &  $ 18.2  $ & $-16.3  $ & $ 5.9  $ & $8.6   $ &$ -18.8$&$5.6 $&$ 9.1$\\
			& $\gamma_2 $ & $ -87.5  $  & $ 2.6  $ & $ 79.2  $ & $-90.4 $  & $2.4   $ & $84.0  $ & $ -38.3 $ &$ 7.1$&$ 21.8$&$ -43.1$&$ 6.4$&$ 24.9$\\
			& $\beta $ & $ -48.9  $ & $ 0.4$ & $ 24.3 $ & $ -48.8  $ & $  0.4  $ & $ 24.2$ & $ -24.1 $ & $ 1.1  $ & $  6.9$ & $  -24.0 $& $ 1.1 $ &$6.9 $\\
			& & & & & & & & & && &&\\[-8pt]
			$3/1/2$  & $\gamma_1 $ & $-35.3$  & $4.8$ & $ 17.3$ & $-38.1  $ & $4.9 $ &  $  19.4 $ & $-14.0  $ & $ 8.7  $ & $ 10.7  $ &$ -17.6$&$10.0 $&$ 13.1$\\
			& $\gamma_2 $ & $ -87.6  $  & $ 3.7  $ & $  80.4 $ & $ -90.5$  & $  3.9 $ & $ 85.8 $ & $ -37.8 $ &$10.3 $&$24.6 $&$-43.0 $&$11.8 $&$ 30.4$\\
			& $\beta $ & $ -47.5  $ & $ 0.5$ & $ 23.0 $ & $ -47.4  $ & $ 0.5   $ & $23.0 $ & $-22.6  $ & $1.3   $ & $6.4  $ & $ -22.5  $& $ 1.2 $ &$6.3 $\\
			& & & & & & & & & && &&\\[-8pt]
			$3/2/1$  & $\gamma_1 $ & $-1.1 $  & $1.7$ & $ 1.7$ & $ -1.5 $ & $ 1.6$ &  $1.6   $ & $-1.0  $ & $ 2.2  $ & $ 2.2  $ &$ -1.7$&$2.1 $&$2.1 $\\
			& $\gamma_2 $ & $  -90.5 $  & $  2.0 $ & $ 83.9  $ & $ -93.7$  & $ 1.9  $ & $ 89.8 $ & $ -41.2 $ &$ 5.5$&$ 22.4$&$ -46.3$&$5.3 $&$ 26.7$\\
			& $\beta $ & $ -51.5  $ & $0.7 $ & $ 27.2 $ & $ -51.3  $ & $  0.7  $ & $27.1 $ & $-26.6  $ & $  2.1 $ & $ 9.2 $ & $ -26.5  $& $  2.1$ &$9.1 $\\
			& & & & & & & & & && &&\\[-8pt]
			$3/2/2$  & $\gamma_1 $ & $0.0 $  & $2.7$ & $ 2.7$ & $ -1.7 $ & $ 2.5$ &  $ 2.5  $ & $0.1  $ & $  3.5 $ & $ 3.5  $ &$ -1.5$&$ 3.6$&$3.6 $\\
			& $\gamma_2 $ & $ -91.0  $  & $  3.2 $ & $  85.9 $ & $-95.4 $  & $ 3.0  $ & $ 94.1 $ & $ -41.9 $ &$8.9 $&$26.5 $&$ -49.1$&$ 8.7$&$ 32.9$\\
			& $\beta $ & $ -49.9  $ & $ 1.2$ & $26.1  $ & $  -49.5 $ & $  1.2  $ & $25.7 $ & $-25.1  $ & $ 3.1  $ & $ 9.4 $ & $-24.6   $& $ 3.0 $ &$9.0 $\\
	\end{tabular}}
	\caption{Bias, variance and MSE of $\hat\gamma$ and $\hat\beta$ for the naive and simex method using the maximum likelihood (1) and the  presmoothing (2) approach for Model 1 ($n=400$). The first column gives the setting/scenario/cens. level. All numbers were multiplied by $100$.}
	\label{tab:results1_400}
\end{table}

\begin{table}[h]
	\centering
	\scalebox{0.85}{
		\begin{tabular}{ccrrrrrrrrrrrr}
			&&\multicolumn{3}{c}{naive - 1}&\multicolumn{3}{c}{naive - 2}&\multicolumn{3}{c}{simex - 1}&\multicolumn{3}{c}{simex - 2}\\
			Mod./Scen./$v$ 	& Par. &  Bias & Var. & MSE & Bias & Var. & MSE & Bias & Var. & MSE&Bias & Var. & MSE\\[2pt]
				& & & & & & & & & && &&\\[-8pt]
			$2/2/0.2 $ & $\gamma_1 $ & $2.5 $  & $6.8 $ & $6.9  $ & $1.9  $ & $6.8 $ & $6.8 $ & $ 4.0 $ & $ 7.1$ & $7.2 $ & $3.7 $ & $7.5 $ & $ 7.7$\\
			& $\gamma_2 $ & $-11.7 $  & $ 10.2$ & $ 11.6 $ & $ -13.9 $ & $9.8 $ & $ 11.8$ & $ 2.0 $ & $13.4 $ & $ 13.4$ & $-0.3 $ & $13.9 $ & $ 14.0$\\
			& $\gamma_3$ & $-2.8 $  & $12.7 $ & $ 12.8 $ & $  -2.9$ & $ 12.7$ & $12.8 $ & $-3.0  $ & $ 13.0$ & $ 13.1$ & $ -3.3$ & $13.9 $ & $14.0 $\\
			& $\beta_1 $ & $-33.3 $  & $ 4.1$ & $  15.2$ & $ -33.3 $ & $ 4.1$ & $15.2 $ & $ -1.4 $ & $ 7.9$ & $ 7.9$ & $ -1.4$ & $ 7.9$ & $7.9 $\\
			& $\beta_2 $ & $6.1 $  & $ 4.3$ & $ 4.7 $ & $ 6.1 $ & $ 4.3$ & $ 4.7$ & $ 0.0 $ & $5.3 $ & $5.3 $ & $ 0.0$ & $5.3 $ & $5.3 $\\
			& & & & & & & & & && &&\\[-8pt]
			$2/2/0.4 $ & $\gamma_1 $ & $ -0.7$  & $ 6.5$ & $ 6.5 $ & $ -1.4 $ & $ 6.6$ & $6.6 $ & $ 2.2 $ & $ 7.2$ & $ 7.2$ & $ 1.7$ & $8.0 $ & $8.0 $\\
			& $\gamma_2 $ & $ -41.8$  & $7.4 $ & $ 24.9 $ & $ -42.9 $ & $7.3 $ & $25.7 $ & $ -10.7 $ & $ 15.0$ & $16.2 $ & $ -11.5$ & $16.4 $ & $17.7 $\\
			& $\gamma_3$ & $-1.7 $  & $12.4 $ & $ 12.4 $ & $ -2.1 $ & $ 12.4$ & $12.4 $ & $ -2.3 $ & $13.2 $ & $ 13.3$ & $-3.0 $ & $14.5 $ & $14.6 $\\
			& $\beta_1 $ & $-89.9 $  & $2.5 $ & $ 83.3 $ & $ -89.9 $ & $ 2.5$ & $ 83.3$ & $ -41.1 $ & $7.3 $ & $24.2 $ & $-41.1 $ & $ 7.3$ & $24.2 $\\
			& $\beta_2 $ & $16.8 $  & $ 4.5$ & $ 7.4 $ & $ 16.9 $ & $ 4.5$ & $7.4 $ & $ 8.1 $ & $6.5 $ & $7.1 $ & $8.1 $ & $ 6.5$ & $ 7.1$\\
				& & & & & & & & & && &&\\[-8pt]
	$2/3/0.2 $ & $\gamma_1 $ & $0.0 $  & $7.3 $ & $  7.3$ & $0.0  $ & $7.3 $ & $7.3 $ & $ -0.9 $ & $ 7.7$ & $7.7 $ & $-0.6 $ & $8.0 $ & $8.0 $\\
& $\gamma_2 $ & $-10.3 $  & $12.7 $ & $  13.8$ & $ -16.7 $ & $ 12.0$ & $14.8 $ & $5.4  $ & $16.7 $ & $ 17.0$ & $-1.3 $ & $ 18.0$ & $ 18.0$\\
& $\gamma_3$ & $ 0.5$  & $ 14.4$ & $ 14.4 $ & $-0.7  $ & $14.3 $ & $ 14.3$ & $ 2.5 $ & $ 15.2$ & $ 15.3$ & $1.1 $ & $ 16.1$ & $16.1 $\\
& $\beta_1 $ & $ -8.9$  & $6.3 $ & $ 7.0 $ & $ -8.4 $ & $ 6.2$ & $6.9 $ & $ 1.5 $ & $9.0 $ & $9.0 $ & $2.0 $ & $8.9 $ & $ 9.0$\\
& $\beta_2 $ & $-0.6 $  & $ 7.2$ & $ 7.2 $ & $ -0.6 $ & $7.2 $ & $7.2 $ & $ 0.7 $ & $7.6 $ & $ 7.6$ & $0.7 $ & $ 7.5$ & $ 7.6$\\
& & & & & & & & & && &&\\[-8pt]
$2/3/0.4 $ & $\gamma_1 $ & $1.7 $  & $ 6.8$ & $ 6.8 $ & $ 1.9 $ & $ 7.0$ & $7.1 $ & $-0.1  $ & $ 7.6$ & $ 7.6$ & $0.9 $ & $9.0 $ & $9.0 $\\
& $\gamma_2 $ & $-45.2 $  & $9.2 $ & $29.6  $ & $ -51.5 $ & $8.4 $ & $35.0 $ & $-8.0  $ & $19.5 $ & $ 20.1$ & $ -16.5$ & $ 19.8$ & $22.5 $\\
& $\gamma_3$ & $-3.2 $  & $13.3 $ & $13.4  $ & $ -4.6 $ & $ 13.5$ & $ 13.7$ & $  0.9$ & $ 15.1$ & $ 15.1$ & $ -1.3$ & $ 17.0$ & $17.1 $\\
& $\beta_1 $ & $ -29.2$  & $ 4.2$ & $ 12.8 $ & $  -28.8$ & $4.2 $ & $ 12.5$ & $- 9.5 $ & $ 9.3$ & $ 10.2$ & $ -8.9$ & $9.3 $ & $10.0 $\\
& $\beta_2 $ & $-3.1 $  & $ 7.2$ & $ 7.3 $ & $ -3.1 $ & $ 7.2$ & $7.3 $ & $ -0.6 $ & $8.0 $ & $8.0 $ & $-0.5 $ & $8.0 $ & $ 8.0$\\
	& & & & & & & & & && &&\\[-8pt]
$3/1/0.1 $ & $\gamma_1 $  & $  5.4$ & $11.2 $ & $11.5 $ & $4.2$  & $11.3 $ & $ 11.5 $& $ 5.4 $ & $ 11.2$ & $11.5 $ & $4.2 $ & $11.3 $ & $11.5 $\\
& $\gamma_2 $  & $ 3.0 $ & $20.4 $ & $ 20.5$& $-0.1$  & $ 20.8$ & $20.8  $ & $ 3.0 $ & $10.4 $ & $20.5 $ & $ -0.1$ & $ 20.8$ & $ 20.8$\\
& $\gamma_3$  & $ 0.5 $ & $ 24.9$&  $ 24.9$ & $ -1.0$  & $ 26.3$ & $ 26.3 $ & $ 0.5 $ & $24.9 $ & $24.9 $ & $-1.0 $ & $26.3 $ & $ 26.3$\\
& $\beta_1 $ & $0.9$  & $ 4.3$ & $ 4.3 $ & $ 1.0 $ & $ 4.3$ & $ 4.3$ & $  1.1$ & $ 4.3$ & $ 4.3$ & $ 1.1$ & $4.3 $ & $ 4.3$\\
& $\beta_2 $ & $ -5.8$  & $ 9.6$ & $9.9  $ & $ -6.0 $ & $9.5 $ & $9.9 $ & $  -0.9$ & $12.1 $ & $ 12.1$ & $ 1.1$ & $12.1 $ & $ 12.1$\\
& & & & & & & & & && &&\\[-8pt]
$2/1/0.2 $ & $\gamma_1 $ & $5.4 $  & $ 11.2$ & $ 11.4 $ & $4.2  $ & $ 11.3$ & $11.5 $ & $ 5.4 $ & $11.2 $ & $ 11.4$ & $4.2 $ & $11.3 $ & $11.5 $\\
& $\gamma_2 $ & $2.8 $  & $20.4 $ & $ 20.4 $ & $-0.2  $ & $20.9 $ & $ 20.9$ & $ 2.9 $ & $20.4 $ & $20.5 $ & $ -0.2$ & $20.9 $ & $20.9 $\\
& $\gamma_3$ & $ 0.4$  & $ 24.8$ & $ 24.8 $ & $ -1.0 $ & $26.2 $ & $26.2 $ & $  0.3$ & $24.8 $ & $ 24.8$ & $ -1.0$ & $26.2 $ & $26.2 $\\
& $\beta_1 $ & $0.6 $  & $4.3 $ & $4.3  $ & $ 0.6 $ & $4.2 $ & $ 4.2$ & $ 0.9 $ & $4.4 $ & $ 4.4$ & $ 0.9$ & $ 4.3$ & $4.3 $\\
& $\beta_2 $ & $-16.6 $  & $7.2 $ & $9.9  $ & $-16.7  $ & $ 7.2$ & $ 10.0$ & $-5.6  $ & $13.4 $ & $ 13.7$ & $-5.8 $ & $13.3 $ & $13.6 $\\
		& & & & & & & & & && &&\\[-8pt]
$3/2/0.1 $ & $\gamma_1 $ & $4.4 $  & $7.3 $ & $ 7.5 $ & $ 3.6 $ & $7.2 $ & $ 7.4$ & $ 4.3 $ & $ 7.3$ & $ 7.5$ & $ 3.6$ & $ 7.2$ & $7.4 $\\
& $\gamma_2 $ & $3.3 $  & $12.0 $ & $12.1  $ & $-0.5  $ & $11.8 $ & $11.8 $ & $ 3.3 $ & $ 12.0$ & $12.1 $ & $-0.5 $ & $11.8 $ & $11.8 $\\
& $\gamma_3$ & $-2.5 $  & $12.5 $ & $ 12.5 $ & $ -2.4 $ & $ 12.3$ & $ 12.3$ & $-2.5  $ & $12.5 $ & $12.5 $ & $-2.4 $ & $12.3 $ & $ 12.3$\\
& $\beta_1 $ & $0.5 $  & $3.7 $ & $ 3.7 $ & $ 0.6 $ & $3.7 $ & $3.7 $ & $ 1.1 $ & $3.7 $ & $3.7 $ & $1.2 $ & $ 3.7$ & $ 3.7$\\
& $\beta_2 $ & $8.3 $  & $ 9.0$ & $ 9.7 $ & $ 8.3 $ & $ 9.0$ & $ 9.7$ & $ -1.8 $ & $11.6 $ & $11.6 $ & $-1.8 $ & $11.6 $ & $ 11.6$\\
& & & & & & & & & && &&\\[-8pt]
$3/2/0.2 $ & $\gamma_1 $ & $4.3 $  & $ 7.3$ & $ 7.5 $ & $ 3.6 $ & $7.2 $ & $ 7.4$ & $ 4.3 $ & $7.3 $ & $ 7.5$ & $ 3.6$ & $ 7.2$ & $7.4 $\\
& $\gamma_2 $ & $3.3 $  & $12.0$ & $ 12.1 $ & $-0.5  $ & $11.8 $ & $ 11.8$ & $  3.3$ & $12.0 $ & $12.1 $ & $-0.5 $ & $11.8 $ & $ 11.8$\\
& $\gamma_3$ & $ -2.5$  & $ 12.4$ & $ 12.5 $ & $  -2.4$ & $ 12.3$ & $12.3 $ & $ -2.5 $ & $12.5 $ & $12.1 $ & $-2.4 $ & $ 12.3$ & $12.3 $\\
& $\beta_1 $ & $-0.8 $  & $3.7$ & $ 3.7$ & $ -0.7 $ & $ 3.7$ & $3.7 $ & $ 0.5 $ & $3.9 $ & $ 3.9$ & $0.6 $ & $3.9 $ & $ 3.9$\\
& $\beta_2 $ & $30.4$  & $6.8 $ & $ 16.1$ & $ 30.4 $ & $ 6.8$ & $ 16.1$ & $ 7.3 $ & $ 13.1$ & $ 13.6$ & $ 7.3$ & $13.0 $ & $ 13.6$\\
\end{tabular}}
\caption{Bias, variance and MSE of $\hat\gamma$ and $\hat\beta$ for the naive and simex method  based on the maximum likelihood  (1) or the  presmoothing  (2) approach in Models 2 and 3 ($n=200$). The first column gives the model, scenario and the standard deviation of the measurement error. All numbers were multiplied by $100$.}
\label{tab:results_3}
\end{table}

\begin{table}[h]
	\centering
	\scalebox{0.85}{
		\begin{tabular}{ccrrrrrrrrrrrr}
			& & & & & & & & && &&&\\[-10pt]
			&&\multicolumn{3}{c}{naive - 1}&\multicolumn{3}{c}{naive - 2}&\multicolumn{3}{c}{simex - 1}&\multicolumn{3}{c}{simex - 2}\\
			Mod./Scen./$v$ 	& Par. &  Bias & Var. & MSE & Bias & Var. & MSE & Bias & Var. & MSE&Bias & Var. & MSE\\[2pt]
			& & & & & & & & & && &&\\[-8pt]
		$4/1$  & $\gamma_1 $ & $3.4 $  & $ 5.3$ & $  5.4$ & $ 1.6 $ & $ 5.1$ & $ 5.2$ & $4.3  $ & $5.5 $ & $5.7 $ & $2.4 $ & $5.6 $ & $5.6 $\\
		$v_1=0.35$ & $\gamma_2 $ & $ -2.2$  & $ 4.9$ & $4.9  $ & $ -5.1 $ & $ 4.5$ & $4.8 $ & $3.8  $ & $ 6.4$ & $6.5 $ & $0.2 $ & $6.7 $ & $6.7 $\\
		$v_2=0.2$	& $\beta_1$ & $ -6.5$  & $1.0 $ & $ 1.5 $ & $ -6.3 $ & $1.0 $ & $1.4 $ & $ -0.5 $ & $1.4 $ & $1.4 $ & $ -0.3$ & $1.4 $ & $1.4 $\\
		& $\beta_2 $ & $0.1 $  & $ 2.6$ & $ 2.6 $ & $0.1  $ & $ 2.6$ & $ 2.6$ & $ 1.5 $ & $3.4 $ & $3.5 $ & $1.5 $ & $ 3.4$ & $ 3.4$\\
		& & & & & & & & & && &&\\[-8pt]
		$4/1$ & $\gamma_1 $ & $ 1.8$  & $5.2 $ & $ 5.3 $ & $ 0.1 $ & $5.2 $ & $5.2 $ & $ 3.6 $ & $5.7 $ & $5.8 $ & $2.0 $ & $ 6.2$ & $6.2 $\\
		$v_1=0.7$& $\gamma_2 $ & $-14.3 $  & $3.6 $ & $ 5.6 $ & $ -16.7 $ & $3.4 $ & $ 6.2$ & $-1.4  $ & $7.0 $ & $7.0 $ & $-4.8 $ & $7.4 $ & $ 7.7$\\
		$v_2=0.4$&  $\beta_1 $ & $ -18.8$  & $0.8 $ & $ 4.3 $ & $ -18.7 $ & $0.8 $ & $ 4.3$ & $ -6.8 $ & $ 1.7$ & $ 2.1$ & $ -6.7$ & $1.7 $ & $2.1 $\\
		& $\beta_2 $ & $ -2.7$  & $ 2.0$ & $2.1  $ & $ -2.7 $ & $ 2.0$ & $2.1 $ & $ 0.0 $ & $ 3.9$ & $3.9 $ & $ 0.0$ & $ 3.9$ & $3.9 $\\
		& & & & & & & & & && &&\\[-8pt]
		$4/3$ & $\gamma_1 $ & $0.7 $  & $6.1 $ & $6.1  $ & $ -1.2 $ & $ 5.9$ & $ 5.9$ & $ 0.3 $ & $7.1 $ & $ 7.1$ & $-1.4 $ & $7.4 $ & $7.4 $\\
		$v_1=0.35$ & $\gamma_2 $ & $ 29.4$  & $9.9 $ & $ 18.6 $ & $  37.7$ & $10.5 $ & $ 24.8$ & $  -3.7$ & $18.3 $ & $ 18.4$ & $ 8.2$ & $20.8 $ & $21.5 $\\
		$v_2=0.2$&  $\beta_1 $ & $33.6 $  & $ 3.9$ & $ 15.2 $ & $ 33.3 $ & $4.0 $ & $ 15.0$ & $ 3.7 $ & $8.2 $ & $ 8.3$ & $3.2 $ & $ 8.2$ & $ 8.3$\\
		& $\beta_2 $ & $ -9.0$  & $ 4.8$ & $5.6  $ & $-9.3  $ & $4.7 $ & $5.6 $ & $ -0.8 $ & $7.6 $ & $ 7.6$ & $-1.0 $ & $ 7.5$ & $ 7.5$\\
		& & & & & & & & & && &&\\[-8pt]
		$4/3$ & $\gamma_1 $ & $1.2 $  & $5.2 $ & $ 5.2 $ & $ -0.9 $ & $4.9 $ & $5.0 $ & $ 0.9 $ & $6.9 $ & $6.9 $ & $-1.0 $ & $6.9 $ & $7.0 $\\
		$v_1=0.7$	& $\gamma_2 $ & $89.1 $  & $ 4.9$ & $ 84.3 $ & $ 93.9 $ & $ 5.0$ & $ 93.1$ & $ 38.8 $ & $ 13.8$ & $ 28.9$ & $ 46.6$ & $ 14.7$ & $36.4 $\\
		$v_2=0.4$&  $\beta_1 $ & $82.2 $  & $2.6 $ & $ 70.2 $ & $ 82.0 $ & $ 2.5$ & $ 69.8$ & $ 46.2 $ & $ 7.8$ & $29.2 $ & $45.9 $ & $7.6 $ & $28.7 $\\
		& $\beta_2 $ & $-23.1 $  & $3.7 $ & $ 9.0 $ & $ -23.3 $ & $ 3.6$ & $ 9.0$ & $  -11.2$ & $8.7 $ & $10.0 $ & $- 11.5$ & $ 8.5$ & $9.9 $\\
		& & & & & & & & & && &&\\[-8pt]
		$5/1/0.39 $ & $\gamma_1 $ & $3.2 $  & $ 4.9$ & $5.0  $ & $ 1.2 $ & $4.7 $ & $ 4.7$ & $3.2  $ & $ 5.0$ & $ 5.1$ & $ 1.2$ & $4.7 $ & $4.7 $\\
		& $\gamma_2 $ & $2.9 $  & $5.5 $ & $ 5.6 $ & $-0.5  $ & $ 5.3$ & $5.3 $ & $  2.9$ & $5.6 $ & $5.6 $ & $-0.5 $ & $5.3 $ & $5.3 $\\
		& $\beta_1 $ & $ -3.7$  & $ 3.3$ & $ 3.4 $ & $ -3.5 $ & $ 3.2$ & $3.4 $ & $ -1.5 $ & $5.7 $ & $ 5.7$ & $ -1.3$ & $5.6 $ & $5.6 $\\
		& $\beta_2 $ & $ -4.3$  & $2.4 $ & $2.5  $ & $ -4.3 $ & $ 2.3$ & $2.5 $ & $ -2.0 $ & $ 4.7$ & $ 4.8$ & $-2.1 $ & $ 4.7$ & $4.8 $\\
		& & & & & & & & & && &&\\[-8pt]
		$5/1/0.78 $ & $\gamma_1 $ & $3.1 $  & $4.9 $ & $ 5.0 $ & $ 1.2 $ & $ 4.7$ & $ 4.7$ & $ 3.2 $ & $ 5.0$ & $ 5.1$ & $1.2 $ & $4.7 $ & $4.7 $\\
		& $\gamma_2 $ & $ 2.9$  & $ 5.5$ & $ 5.6 $ & $ -0.5 $ & $ 5.3$ & $5.3 $ & $ 2.9 $ & $5.6 $ & $5.6 $ & $ -0.5$ & $5.3 $ & $5.3 $\\
		& $\beta_1 $ & $-6.7 $  & $ 2.1$ & $2.5  $ & $ -6.5 $ & $2.0 $ & $2.5 $ & $ -5.0 $ & $4.1 $ & $4.3 $ & $-4.8 $ & $4.0 $ & $ 4.3$\\
		& $\beta_2 $ & $-7.2 $  & $ 1.1$ & $ 1.6 $ & $ -7.3 $ & $ 1.1$ & $1.6 $ & $ -5.4 $ & $ 3.1$ & $3.4 $ & $-5.5 $ & $3.1 $ & $ 3.4$\\
		& & & & & & & & & && &&\\[-8pt]
		$5/2/0.39 $ & $\gamma_1 $ & $ 6.1$  & $ 7.9$ & $ 8.3 $ & $2.2  $ & $7.1 $ & $ 7.2$ & $ 6.1 $ & $7.9 $ & $8.3 $ & $ 2.2$ & $ 7.1$ & $7.2 $\\
		& $\gamma_2 $ & $ 8.6$  & $14.5 $ & $ 15.2 $ & $ -1.8 $ & $ 12.5$ & $12.6 $ & $ 8.6 $ & $ 14.5$ & $ 15.2$ & $ -1.8$ & $12.5 $ & $12.6 $\\
		& $\beta_1 $ & $ 18.4$  & $ 3.2$ & $  6.6$ & $18.5  $ & $3.2 $ & $6.6 $ & $ 6.6 $ & $5.2 $ & $5.7 $ & $ 6.7$ & $5.2 $ & $5.7 $\\
		& $\beta_2 $ & $ 18.6$  & $ 2.2$ & $ 5.6 $ & $ 18.6 $ & $ 2.2$ & $5.6 $ & $ 6.1 $ & $ 4.5$ & $ 4.9$ & $6.2 $ & $4.5 $ & $4.9 $\\
		& & & & & & & & & && &&\\[-8pt]
		$5/2/0.78 $ & $\gamma_1 $ & $6.2 $  & $ 8.0$ & $ 8.3 $ & $ 2.2 $ & $ 7.1$ & $7.2 $ & $6.1  $ & $8.0 $ & $ 8.4$ & $ 2.2$ & $ 71.$ & $7.2 $\\
		& $\gamma_2 $ & $ 8.7$  & $14.5 $ & $ 15.3 $ & $ -1.8 $ & $ 12.5$ & $ 12.6$ & $8.6  $ & $ 14.6$ & $ 15.3$ & $-1.8 $ & $ 12.5$ & $12.6 $\\
		& $\beta_1 $ & $34.3 $  & $2.3 $ & $  14.0$ & $  34.4$ & $ 2.3$ & $14.1 $ & $ 25.3 $ & $ 3.9$ & $10.3 $ & $25.4 $ & $ 3.8$ & $ 10.3$\\
		& $\beta_2 $ & $35.5 $  & $ 1.0$ & $ 13.6 $ & $35.5  $ & $ 1.0$ & $13.6 $ & $26.0  $ & $ 2.8$ & $ 9.6$ & $26.0 $ & $ 2.8$ & $ 9.6$\\		
			\end{tabular}}
	\caption{Bias, variance and MSE of $\hat\gamma$ and $\hat\beta$ for the naive and simex method  based on the maximum likelihood  (1) or the  presmoothing  (2) approach in Models 4 and 5 ($n=200$). The first column gives the model, scenario and the standard deviation of the measurement error. All numbers were multiplied by $100$.}
	\label{tab:results_4}
\end{table}

%

\end{document}